\documentclass[superscriptaddress,nobibnotes,amsmath,amssymb,notitlepage,twocolumn,prl,longbibliography]{revtex4-2}
\usepackage[utf8]{inputenc}

\usepackage{bm,mathptmx,braket}

\usepackage{bbm}
\usepackage{graphicx}
\usepackage[dvipsnames]{xcolor}
\usepackage[colorlinks=true, linkcolor=blue, citecolor=blue, urlcolor=blue]{hyperref}
\usepackage[caption=false]{subfig}

\newcommand{\sect}[1]{\vspace{0.3em}{\it #1.}---}

\graphicspath{{./img/}}

\newcommand{\beq}[1]{\begin{equation}\label{#1}}
\newcommand{\eep}{\;.\end{equation}}
\newcommand{\eec}{\;,\end{equation}}
\newcommand{\eeq}{\end{equation}}

\newcommand*\dd{\mathop{}\!\mathrm{d}} 

\newcommand{\lb}{\left(}
\newcommand{\rb}{\right)}

\newcommand{\om}{\omega}

\DeclareMathAlphabet{\mathcal}{OMS}{cmsy}{m}{n} 

\renewcommand{\vec}[1]{{\bf #1}}

\newcommand{\kv}{\vec{k}}

\begin{document}

\title{Quantum geometric bounds in non-Hermitian systems}

\date{\today}

\newcommand{\TCM}{{TCM Group, Cavendish Laboratory, University of Cambridge, J.\,J.\,Thomson Avenue, Cambridge CB3 0HE, UK}}
\newcommand{\MPIPKS}{{Max-Planck-Institut f\"ur Physik komplexer Systeme, N\"othnitzer Stra\ss e 38, 01187 Dresden, Germany}}
\newcommand{\quantinuum}{Quantinuum, Terrington House, 13-15 Hills Rd, Cambridge, CB2 1NL, UK}

\author{Milosz Matraszek}
\email{mpm70@cam.ac.uk}
\affiliation{\TCM}
\affiliation{\MPIPKS}

\author{Wojciech J. Jankowski}
\email{wjj25@cam.ac.uk}
\affiliation{\TCM}
\affiliation{\MPIPKS}

\author{Jan Behrends}
\email{jan.behrends@yahoo.de}
\altaffiliation[Present address: ]{\quantinuum}
\affiliation{\TCM}

\begin{abstract}
    We identify quantum geometric bounds for observables in non-Hermitian systems. We find unique bounds on non-Hermitian quantum geometric tensors, generalized two-point response correlators, conductivity tensors, and optical weights. We showcase these findings in topological systems with non-Hermitian Chern numbers. We demonstrate that the non-Hermitian geometric constraints on response functions naturally arise in open quantum systems governed by out-of-equilibrium Lindbladian dynamics. Our findings are relevant to experimental observables and responses under the realistic setups that fall beyond the idealized closed-system descriptions.
\end{abstract}

\maketitle

\sect{Introduction} Geometry plays a pivotal role in physics, from classical dynamics and general relativity, to statistical thermodynamics and quantum mechanics. While symplectic geometry underpins the Hamiltonian formulation of theoretical mechanics and analytical thermodynamics~\cite{analytical_mechanics}, Riemannian geometry captures multiple physical features in gravitational~\cite{o1983semi} and quantum physics~\cite{Provost1980}. In both latter fields, multiple physical properties are efficiently captured by the metric tensors and their derivatives.

In quantum mechanics, the Riemannian quantum metric, introduced by Provost and Valle\'e~\cite{Provost1980}, encodes the measure of distance between distinct quantum states in projective Hilbert spaces, and is constituent to versatile quantum geometric tensors (QGTs)~\cite{Anandan:1990dx,Kolodrubetz:2017jg,Torma2023,Verma2025}. Physical manifestations of quantum distance and geometry range from the realm of quantum metrology underpinned by quantum Fisher information and quantum Cram\'er--Rao bounds~\cite{Facchi:2010hl,Liu2019, Yu2022, Bouhon2023, Yu2024a, Wahl2025, Jankowski2025PRR} to solid state materials~\cite{Marzari1997, Rhim2020}. In condensed matter systems, the quantum geometry is of key importance to correlated phenomena such as flat-band superconductivity~\cite{Peotta2015, Xie2020, Verma2021, Peri2021a, HerzogArbeitman2022,Torma:2022jg} or excitonic physics~\cite{Mao2024, Hu:2024kw, Jankowski2025exciton, Thompson2024exciton}. Nontrivial quantum geometries are manifested in electric transport and optical phenomena of free and interacting particles, coupled to light at linear and nonlinear orders~\cite{Ahn2020, Ahn2021, Onishi:2024gl, Jankowski2024PRL, Yu2025, Jankowski2025PRL}. 

Remarkably, recent experimental advances involve resolving the quantum metric and QGTs over the space of electron momenta in solid state materials~\cite{Kang2025, Kim2025}. Crucially, QGTs and quantum metric provide local bounds on physical observables, owing to positive semidefiniteness conditions~\cite{Bouhon2023, Shinada2025}. Particularly interesting are the bounds due to the nontrivial topologies of quantum states~\cite{Peotta2015, Xie2020, Jankowski2025PRBoptical, Jankowski2025exciton, Jankowski2025PRL}. A general limitation is that these results on quantum geometry consider only closed systems governed by Hermitian Hamiltonians and thermal density matrices~\cite{Shinada2025, Ji2025}.
Quantum geometry and bounds in driven-dissipative systems have been considered only recently for slow external driving~\cite{Esin2025}, and for the localization of Wannier functions~\cite{Montag2025}.

Quantum geometric quantities have recently been extended to non-Hermitian (NH) systems~\cite{Singhal:2023gl,Ye2024,Ozawa:2025bv,Behrends2025}.
Importantly, such NH systems~\cite{Ashida:2020fo,Bergholtz:2021kc} naturally realize exotic eigenvalue \mbox{spectra} \cite{Herviou:2019ih, Wanjura:2020jg, Borgnia:2020hi, Zirnstein:2021cl} and topologies of quantum states~\cite{Shen2018, Gong:2018ko}.
Furthermore, an~effective description in terms of a non-Hermitian Hamiltonian can be utilized to classify quadratic Lindbladians~\cite{Prosen:2008dw,Prosen:2010hp,Kos:2017ew,Lieu:2020bc,Sa:2023go,Kawabata:2023hk} that govern open quantum systems.
Combining these non-Hermitian paradigms with the notions of nontrivial quantum geometry is an ongoing pursuit~\cite{Behrends2025,Montag2025}.

In NH systems, the nontrivial quantum geometry manifests itself as an anomalous contribution to wave-packet dynamics~\cite{Xu:2017bl,Silberstein:2020hi,Hu:2025ct}.
As a consequence, conductivity in NH systems can be described in terms of geometric quantities: quantum metric and anomalous Berry connection~\cite{Behrends2025}.
However, apart from bounds on the localization of Wannier functions~\cite{Montag2025}, no geometric bounds have hitherto been retrieved.
The fundamental question regarding the existence of physically observable quantum geometric bounds and their manifestation in general NH and open quantum systems has yet to be \mbox{answered}.

In this Letter, we address this question and provide positive answers. {We consider general quantum geometric constraints in the non-Hermitian systems and their responses, which are distinct from the dissipation minimization conditions in slowly-driven open quantum systems of Ref.~\cite{Esin2025}.}
First, we employ the general setups of quantum response theory to show how the quantum geometric bounds persist in the presence of quasiparticle decays in dissipative systems.
We next demonstrate that non-Hermitian QGTs encode distinct bounds from their Hermitian counterparts.
As a consequence, we find that these bounds uniquely constrain the time-dependent quantum responses of topological NH systems. We showcase these non-Hermitian bounds in NH Rice--Mele (RM) models~\cite{Rice:1982cr,Lin:2015jk,Kunst2018} realizing nontrivial NH Chern invariants.
Furthermore, we phrase the retrieved NH quantum geometric bounds as constraints in open quantum systems governed by quadratic Lindbladians~\cite{Prosen:2008dw,Prosen:2010hp,Kos:2017ew}, specifically as constraints of the externally coupled bath.
By employing the Keldysh formalism~\cite{Kamenev_2011,Sieberer:2016ej,Talkington:2024gf,Sieberer:2025eo}, we find that nontrivial geometric bounds arise in bubble diagrams that are intrinsic to response functions of more general open quantum systems.
Notably, these findings are highly relevant to experimental platforms, such as quantum-optical systems or circuit-quantum electrodynamics~\cite{Weimer:2021kp}, where dissipation can be engineered to have a~tailored reservoir.
Our work shows why the geometric bounds intrinsic to response theories hold in typical steady states of externally coupled quantum materials and establishes to what degree these quantum geometric bounds may hold in general NH and open quantum systems.

\sect{Geometry in dissipative quantum responses} We first address the quantum geometric bounds in general dissipative quantum responses from eigenstates $\ket{n}$ dressed by self-energies ${\Sigma}_n = {\Sigma_n'}-\text{i}{\Sigma}_n''$.
We consider a two-point correlator ${\Pi_{O_i O_j}(\om) = \langle O_i(\om) O_j (\om)\rangle}$ for two operators $O_i(\om), O_j(\om)$ in the frequency domain $\omega$, with $i,j$ as the spatial vector components. In the Lehmann representation with a thermal density matrix $\rho$ and the volume of the system $V$~\cite{AltlandSimons2010},
\begin{equation}
    \Pi_{O_i O_j}(\om) = -\frac{1}{V} \sum_{n,m} \rho_{nm} \frac{O^i_{nm} O^j_{mn}}{\om + E_{nm} + \Sigma_{nm} + \text{i}0^+ },
\end{equation}
where $O^i_{nm} = \braket{n|O_i|m}$ are the matrix elements of the operator~$O_i$, ${\Sigma_{nm} \equiv \Sigma_{n}^* - \Sigma_{m}}$, ${\rho_{nm} \equiv \rho_{n} - \rho_{m}}$ with $\rho_{n} = \braket{n|\rho|n}$, and ${E_{nm} \equiv E_{n} - E_{m}}$ are the differences in energies $E_n, E_m$ of eigenstates $\ket{n},\ket{m}$. Upon defining the absorptive part of the correlation function, ${\Pi^{\text{abs}}_{O_i O_j} \equiv \frac{1}{2 \text{i}} [\Pi_{O_i O_j} - (\Pi_{O_j O_i})^*]}$~\cite{Shinada2025}, we obtain geometric bounds if for any complex vector $\textbf{v}$ with coefficients $v_j$,
\begin{equation}\label{eq:posdef}
    v^*_i \Pi^{\text{abs}}_{O_i O_j} v_j = \frac{\pi}{V} \sum_{n,m} \rho_{nm} |O^j_{mn}v_j|^2 L_{nm}(\om) \geqslant 0.
\end{equation}
Here, we define frequency-dependent Lorentzian kernels ${L_{nm}(\om) = \pi^{-1}\Sigma_{nm}''/[(\om+E_{nm}+\Sigma_{nm}')^2 + (\Sigma_{nm}'')^2] \geqslant 0}$, and further recognize that when $E_{nm}<0$, then $\rho_{nm}, \Sigma_{nm}'' \geqslant 0$ for a~dissipative thermal response.
Thus, by Eq.~\eqref{eq:posdef}, $\Pi^{\text{abs}}_{O_i O_j}$ is positive semidefinite. This translates into general nontrivial bounds of the response functions~\cite{Shinada2025},
\begin{equation}
\sqrt{\text{Re}~\Pi^{\text{abs}}_{O_i O_i}\text{Re}~\Pi^{\text{abs}}_{O_j O_j}} \geqslant \Big|\text{Im}~\Pi^{\text{abs}}_{O_i O_j}\Big|.
\end{equation}
%
Notably, if $\Sigma_{nm}'' < 0$, the Lorentzian kernel $L_{nm}(\omega)$ inverts its sign. In that case, $v^*_i \Pi^{\text{abs}}_{O_i O_j} v_j \leqslant 0$, which violates the positive semidefiniteness condition of the quantum geometric bounds.

Having considered general dissipative responses dressed with self-energies, we now focus on NH systems with NH Hamiltonians admitting complex eigenvalues $\varepsilon_{n} \equiv E_n -  \text{i}\Sigma_n''$. We show that the interplay of the NH spectrum  $\varepsilon_{n}$, NH quantum geometry, and NH topology, determine a unique class of NH quantum geometric bounds.

\begin{figure}[t!]
\centering
  \includegraphics[width=\columnwidth]{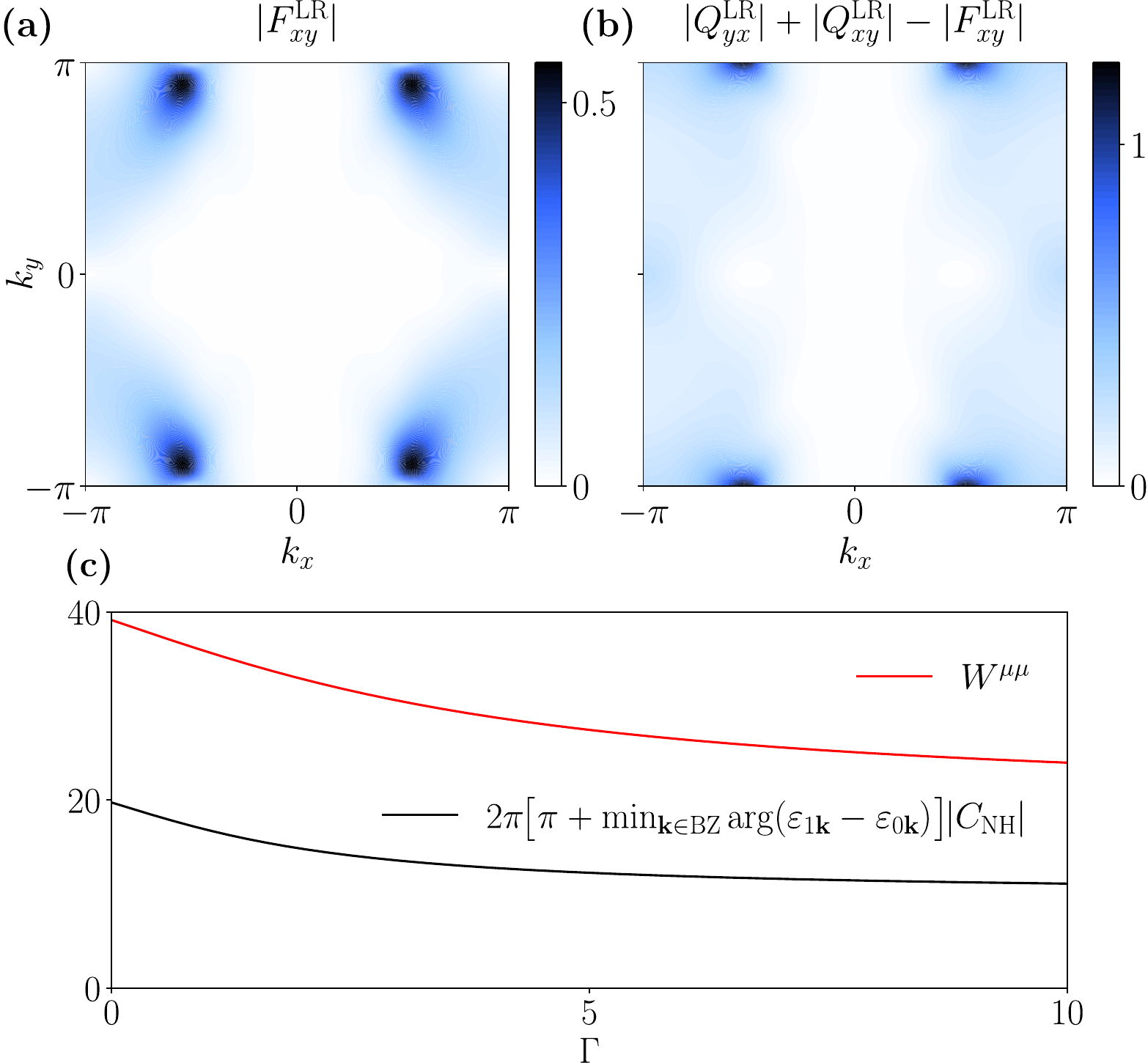}
  \caption{Non-Hermitian quantum geometric bounds. Local Berry curvature bound from NH QGTs in the NH RM model ($\gamma = 1$) [{(a)} and {(b)}]. {(a)} NH Berry curvature over parameter space ($k_x, k_y$). \textbf{(b)} Difference of the NH QGTs and NH Berry curvature, demonstrating that the local bound, Eq.~\eqref{eq:LocalBound}, holds. {(c)} Bound on the optical weights {($W^{\mu\mu}$)} due to NH Chern numbers {($C_{\text{NH}}$)} in the NH RM model.}
\label{fig:NH}
\end{figure}

\sect{Non-Hermitian geometric bounds} In the following, we construct NH geometric bounds from NH integer Chern numbers. Furthermore, we comment on the physical consequences of the bounds due to nonvanishing NH Chern numbers. The NH Chern numbers are defined over a two-dimensional parameter space $\textbf{k} = (k_x, k_y)$ torus that physically can be spanned by momenta of particles or modes in periodic systems~\cite{Shen2018},
\begin{equation}
    C_{\text{NH}} = \frac{1}{2\pi} \int_{T^2} \dd \kv~F^{\text{LR}}_{xy} \in \mathbb{Z}.
\end{equation}
$F^{\text{LR}}_{xy} = \text{i}[\braket{\partial_{k_x}\psi^\text{L}_n | \partial_{k_y}\psi^\text{R}_n} - \braket{\partial_{k_y}\psi^\text{L}_n|\partial_{k_x}\psi^\text{R}_n} ]$ is a NH Berry curvature with $\ket{\psi^\text{L}_n}, \ket{\psi^\text{R}_n}$, the left/right eigenmodes, ${{H} \ket{\psi^\text{R}_n} = \varepsilon_{n\kv} \ket{\psi^\text{R}_n}}$, ${\bra{\psi^\text{L}_n} {H}  = \bra{\psi^\text{L}_n} \varepsilon_{n\kv}}$, of a NH Hamiltonian~${H} \neq {H}^\dagger$. Centrally to this work, we find a local NH bound on NH Berry curvature that reads
\beq{eq:LocalBound}
   \left|F^{\text{LR}}_{\mu\nu}\right| \leqslant \left|Q^{\text{LR}}_{\mu\nu}\right| + \left|Q^{\text{LR}}_{\nu\mu}\right|,
\eeq
where we define the NH QGTs, {$Q^{\text{LR}}_{\mu\nu} =  (Q^{\text{RL}}_{\nu\mu})^{*}$}, as
\beq{eq:LocalBound2}
   Q^{\text{LR}}_{\mu\nu} = \frac{\bra{\partial_{k_\mu} \psi^\text{L}_n} (1 - \ket{\psi^\text{R}_n} \bra{\psi^\text{L}_n}) \ket{\partial_{k_\nu} \psi^\text{R}_n}}{\braket{\psi^\text{L}_n | \psi^\text{L}_n}\braket{\psi^\text{R}_n | \psi^\text{R}_n}}.
\eeq
{\color{black} {Equation~\eqref{eq:LocalBound}} follows directly from the triangle inequality,
\begin{equation}
    \left|F^{\text{LR}}_{\mu\nu}\right| \equiv \left|Q^{\text{LR}}_{\mu\nu} - Q^{\text{LR}}_{\nu\mu}\right| \leqslant \left|Q^{\text{LR}}_{\mu\nu}\right| + \left|Q^{\text{LR}}_{\nu\mu}\right|.
\end{equation}
} We demonstrate that the bound, Eq.~\eqref{eq:LocalBound}, holds locally at every point of the parameter space, see Fig.~\ref{fig:NH}. { We exemplify this bound in a~generalized NH RM model. The NH RM model provides a~minimal realization of the non-Hermitian topology and bulk-boundary correspondence~\cite{Kunst2018}, which naturally invites for further searches towards profound insights into non-Hermitian quantum geometry and topological bounds. The considered NH RM Hamiltonian reads
\begin{equation}\label{eq:NH-RM}
    H_{\text{NH-RM}} = \Bigg( 1+\frac{\text{i} \Gamma}{2\sqrt{\mathbf{d}\cdot\mathbf{d}}} \Bigg) \textbf{d}(\textbf{k}) \cdot \boldsymbol{\sigma},
\end{equation}
with the vector of Pauli matrices $\boldsymbol{\sigma} = (\sigma_x, \sigma_y, \sigma_z)$, and the pseudospin vector
\begin{equation}
    \textbf{d}(\textbf{k}) = \begin{pmatrix}
         t + \delta \cos k_x + [t - \delta \cos k_x] \cos k_y  \\
         [t - \delta \cos k_x] \sin k_y + \frac{\text{i}}{2} \gamma\\
         -\Delta \sin k_x\\
    \end{pmatrix}.
\end{equation}
$\Gamma$ and $\gamma$ parametrize dissipation. Since the Hamiltonian is not Hermitian, its left and right eigenvalues differ, and its energies are generally complex. In~particular, ${H_{\text{NH-RM}}\ket{\psi^{\text{R}}_{n}} = \varepsilon_{n\textbf{k}} \ket{\psi^{\text{R}}_{n}}}$ ($n=0,1$) with the eigenvalues,
\begin{equation}
    \varepsilon_{n\textbf{k}} = (-1)^{n} \Big(\text{i} \frac{\Gamma}{2} +  \sqrt{\textbf{d} \cdot \textbf{d}} \Big).
\end{equation}
The corresponding right eigenvectors $\ket{\psi^{\text{R}}_n}$ can be explicitly written as:
\begin{equation}
    \ket{\psi^{\text{R}}_{0/1}} = \sqrt{\frac{(d_x + \text{i} d_y)(d^*_x - \text{i} d^*_y)}{2|\textbf{d} \cdot \textbf{d}|-2\text{Re}~[\text{i} d_x d_y \pm d_z \sqrt{\textbf{d} \cdot \textbf{d}}]}} \begin{pmatrix}
         d_x -\text{i}d_y \\
         \pm \sqrt{\textbf{d} \cdot \textbf{d}}-d_z\\
         \end{pmatrix}.
\end{equation}
The left eigenvectors $\ket{\psi^\text{L}_n}$, which satisfy the biorthogonality \mbox{relations}, $\braket{\psi^\text{L}_n|\psi^\text{R}_m}= \delta_{nm}$~\cite{Brody:2014jv}, can be constructed as 
\begin{equation}
{\ket{\psi^{\text{L}}_0} = \frac{\ket{\psi^\text{R}_0}-\braket{\psi^\text{R}_1|\psi^\text{R}_0} \ket{\psi^\text{R}_1}/ \braket{\psi^\text{R}_1|\psi^\text{R}_1}}{\braket{\psi^\text{R}_0|\psi^\text{R}_0}-|\braket{\psi^\text{R}_0|\psi^\text{R}_1}|^2/\braket{\psi^\text{R}_1|\psi^\text{R}_1}}},
\end{equation}
{ and analogously for~$\ket{\psi^{\text{L}}_1}$ upon swapping the indices $n=0,1$.}

Importantly, these formulas allow us to directly compute the non-Hermitian Berry curvature, QGTs, and geometric bounds studied in this work. In~particular, the general local bound, Eq.~\eqref{eq:LocalBound}, can be analytically expressed as a geometric condition satisfied by the complex \mbox{$\textbf{d}$-vector} ($\textbf{d} \in \mathbb{C}^3$). We study the non-Hermitian quantum geometric bounds in the topological phase of the NH RM model ($C_\text{NH} = 1$), by setting $t = \delta = \Delta = 1$ in the following.}

Correspondingly, we show the NH Berry curvature $F^\text{LR}_{xy}$ in a~topological non-Hermitian phase ($C_\text{NH} = 1$) of the NH RM model in Fig.~\ref{fig:NH}(a). In Fig.~\ref{fig:NH}(b), we demonstrate how NH QGTs take larger, geometrically lower-bounded values. NH QGTs $|Q^{\text{LR}}_{\mu\nu}|$ vary from $|F^\text{LR}_{xy}|$, saturating the bound to twice larger values at the edge parameters $\textbf{k}$.

Furthermore, we consider the algebra of NH particle current operators, {by setting,} $O_\mu = j_\mu$~\cite{Supplemental}. {The current operators, ${j_\mu = \partial_{k_\mu} H}$, are defined as the momentum-space derivatives of the Hamiltonian $H$~\cite{Behrends2025}, with $\mu = x,y$. As a central result, we find} a~physical two-band bound on NH optical weights~$W^{\mu \nu}$ that are defined with the regular part of the \mbox{conductivity} tensor, ${\sigma^{\text{reg}}_{\mu \nu}(\om) = \Pi_{j_\mu j_\nu}(\om)/\om}$,~as~\cite{Souza:2000cj, Onishi:2024gl}:
\begin{equation}
W^{\mu \nu} \equiv \int^\infty_0~\dd \om~ \frac{\text{Re}~\Pi_{j_\mu j_\nu}(\om)}{\om^2},
\end{equation}
\begin{equation}\label{eq:OpticalWeightBound}
  \frac{1}{2\pi} \sum_{\mu = x,y} W^{\mu\mu} \geqslant \Big[\pi+\text{min}_{\mathbf{k}\in \text{BZ}}\arg(\varepsilon_{1\mathbf{k}}-\varepsilon_{0\mathbf{k}})\Big] |C_{\text{NH}}|.
\end{equation}
{\color{black} Equation~\eqref{eq:OpticalWeightBound} is derived from (i) the non-Hermitian Kubo \mbox{formula}~\cite{Behrends2025} for ${\sigma^{\text{reg}}_{\mu \nu}(\om) = \Pi_{j_\mu j_\nu}(\om)/\om}$, where we evaluate the frequency integral by employing an IR frequency cutoff $(\eta \rightarrow 0)$, and include all single-particle contributions by performing a Brillouin zone (BZ) integral with vanishing terms arising from isolated gauge-invariant tensors. Using the reality of $Q^{\text{RR}}_{\mu \mu}$ tensor components, we perform a leftover integral of 
\begin{equation}
\sum_{\mu = x,y} W^{\mu\mu} = \int_\text{BZ} \text{d}\vec{k}~\left[\pi~\text{tr}~Q^{\text{RR}} + \text{Im}\left(\text{tr}~Q^{\text{RR}}\ln\left(\varepsilon_{1\vec{k}} - \varepsilon_{0\vec{k}} \right) \right)\right],
\end{equation}
to show its dependence on energy is only through $\arg\left(\varepsilon_{1\vec{k}} - \varepsilon_{0\vec{k}} \right)$, which physically corresponds to dissipation. The dissipation leads to the weakening of the bound compared to the Hermitian case, where ${\Sigma''_n = 0}$. We further use (ii) the nonnegativity of ${[\pi+\arg(\varepsilon_{1\mathbf{k}}-\varepsilon_{0\mathbf{k}})]}$, which is implied by the requirement that the branch cut of the complex logarithm cannot intersect the frequency ($\omega$) \mbox{integration} path $\mathbb{R}_+$. That in turn forces $\arg$ to lie in $[-\pi,0]$ in the lower half-plane~\cite{Supplemental}. \mbox{Finally},~we employ (iii) the positive-semidefiniteness of the non-Hermitian quantum geometric tensor $Q^{\text{RR}}_{\mu\nu}$, $v_\mu^*Q^{\text{RR}}_{\mu\nu}v_\nu \geqslant 0$, for any complex vector with coefficients $v_\nu \in \mathbb{C}$, which sets a~bound ${|F^{\text{RR}}_{\mu \nu}| \leqslant Q^{\text{RR}}_{\mu\mu} + Q^{\text{RR}}_{\nu\nu}}$, and a~further bound on the non-Hermitian Chern number ${|C_{\text{NH}}|=|\frac{1}{2\pi}\int_\text{BZ}\text{d}\textbf{k}~F^{\text{RR}}_{\mu \nu}|}$ upon \mbox{summing} all single-particle contributions. For \mbox{detailed} algebraic derivations, see Supplemental Material (SM)~\cite{Supplemental}.} Hence, in~any topological NH phase with ${|C_{\text{NH}}| > 0}$, the NH topology imposes a geometric bound on the time-dependent response functions of the NH ground state. We showcase this bound numerically in the NH RM model [see Fig.~\ref{fig:NH}(c)].

{Finally, we derive an $N$-band generalization of the two-band bound, for multiband non-Hermitian systems with $N > 2$ bands. The generalized optical weight bound reads,
\begin{equation}\label{eq:NbandFinal}
    \frac{1}{2\pi} \sum_{\mu=x,y} W^{\mu\mu} \geqslant \big[\pi+\min_{m>0,\mathbf{k}\in \text{BZ}}\arg(\varepsilon_{m\mathbf{k}}-\varepsilon_{0\mathbf{k}})\Big]|C_\text{NH}|,
\end{equation}
where the non-Hermitian Chern number $C_\text{NH}$ corresponds to the Chern number of the slowest-decaying band $n=0$, i.e., $\Sigma''_{0} < \Sigma''_{m}$, and the band index $m$ runs over all the remaining ${N-1}$ bands. The $N$-band non-Hermitian optical bound is \mbox{derived} from analogous steps to the two-band inequality, as~we \mbox{detail} in the SM~\cite{Supplemental}.
}

\begin{figure}[t!]
\centering
  \includegraphics[width=\columnwidth]{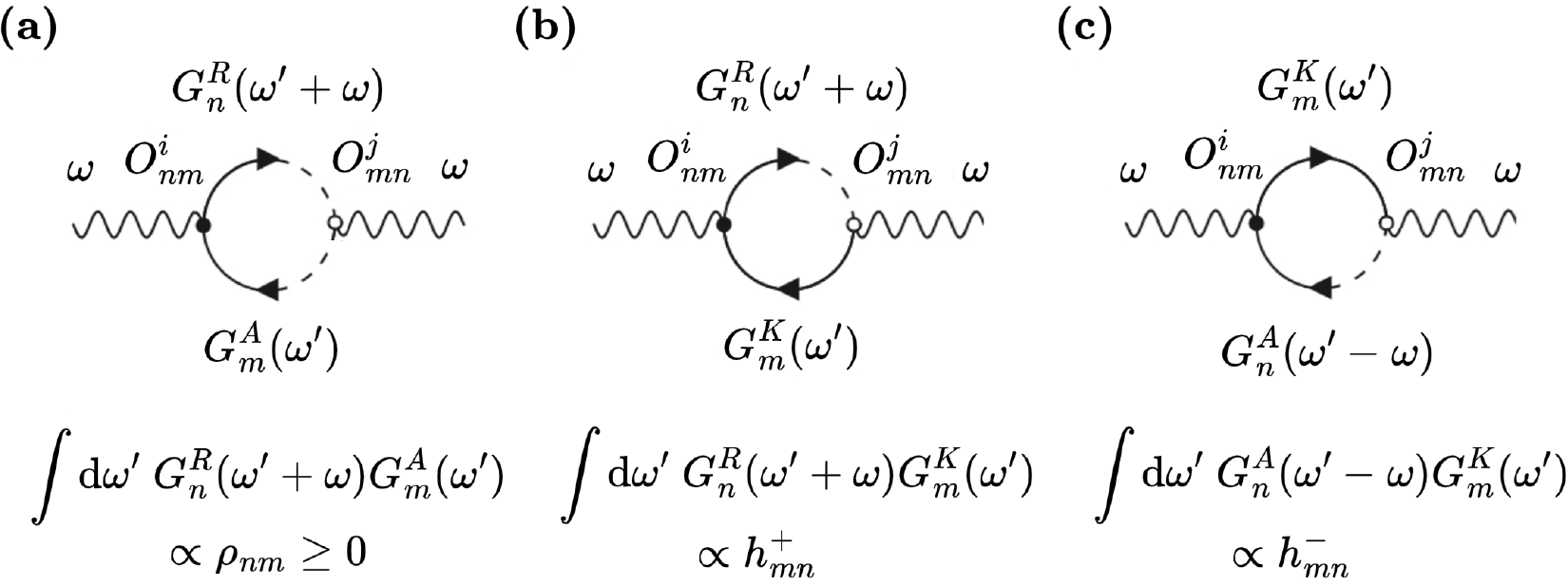}
  \caption{Non-Hermitian geometry from open quantum system \mbox{response} diagrammatics with Keldysh Green's functions. One-loop paramagnetic response diagrams. {(a)} Bubble diagram encoding the geometric bounds in the closed-system limit, {from retarded (${G^R_n}$) and advanced (${G^A_m}$) Green's functions.} {(b)}, {(c)} Geometric \mbox{responses} from Keldysh bubbles upon coupling to a bath inducing the non-Hermiticity. {(b)} Geometric polarization bubble from {retarded} {(${G^R_n}$)} and Keldysh {(${G^K_m}$)} Green's functions. {(c)} Geometric polarization bubble from advanced {(${G^A_n}$)} and Keldysh $({G^K_m})$ Green's functions. The positivity of the {combined} nonequilibrium Green's function sums (${h_{nm}}$) provides for nontrivial geometric bounds in the non-Hermitian systems governed by the open system Lindbladians.}
\label{fig:bubble}
\end{figure}

\sect{Non-Hermitian bounds in open quantum systems} {We~further extend the non-Hermitian quantum geometric bounds to a general setup arising in the responses of open quantum systems within the Lindblad formalism}~\cite{Manzano:2020eb,Talkington:2024gf}.
In particular, any NH Hamiltonian can be mapped to a pair of Lindbladians satisfying quantum geometric bounds in response diagrammatics, as long as the self-energies are diagonal in the Hamiltonian basis~\cite{Supplemental}.
For instance, we can consider a Lindbladian based on the non-Hermitian RM model. {The corresponding matrix representation of the Lindbladian, upon performing a Larkin--Ovchinnikov rotation~\cite{Prosen:2008dw, Talkington:2024gf}, reads},
\begin{equation}
    \mathcal{L} = \begin{pmatrix}
        {H} & \Sigma^K\\
        0 & -{H}^\text{T}
    \end{pmatrix},    
\end{equation}
where $\Sigma^K$ is a Keldysh self-energy. {\color{black} $H$ is an effective non-Hermitian Hamiltonian, which can be written in terms of the Hermitian closed-system Hamiltonian $h(\vec{k})$ and jump \mbox{operators} $J_{\lambda \textbf{k}}$, with ${\lambda = 1,2,\ldots}$,
\begin{equation}
    {H} = {h}(\vec{k}) - \frac{\text{i}}{2}\sum_\lambda J^\dagger_{\lambda\vec{k}}J_{\lambda\vec{k}}.
\end{equation}
Accordingly, $H$ generates the Lindblad master equation, ${\partial_t \rho + \text{i}(H\rho - \rho H^\dagger) = \sum_\lambda J_{\lambda} \rho J^\dagger_{\lambda}}$. The jump operators describing couplings to a bath read
\begin{equation}
    J_{\lambda\vec{k}} \equiv \vec{a}_{\lambda\vec{k}}\cdot \vec{c}_{\vec{k}},
\end{equation}
where $\vec{c}_\kv$ is the vector of annihilation operators. ${\vec{c_\kv}=(c_{0\kv},\,c_{1\kv})}$ for a two-band system. Denoting the bath couplings, ${\vec{a}_{\lambda\vec{k}} = (a_1,\,a_2)}$, the anti-Hermitian part of the non-Hermitian Hamiltonian $M$ is given by
\begin{equation}
    M(a_1,a_2) \equiv \frac{1}{2}(H - H^\dagger) = -\frac{\text{i}}{2}\vec{c^\dagger_\kv} \begin{pmatrix}
        |a_1|^2 & a_1^* a_2 \\
        a_1 a_2^*
        & |a_2|^2
    \end{pmatrix}\vec{c_\kv}.
\end{equation}
Thus, we can construct an arbitrary two-band non-Hermitian matrix from three jump terms $J^\dagger_{\lambda\vec{k}}J_{\lambda\vec{k}}$. Namely, with $\gamma\in\mathbb{R}$, we have
\begin{align}
J_{1\mathbf{k}}^\dagger J_{1\mathbf{k}} &= \gamma\, \mathbf{c}_{\mathbf{k}}^\dagger \sigma^x \mathbf{c}_{\mathbf{k}} + |\gamma|\, \mathbf{c}_{\mathbf{k}}^\dagger \mathbf{c}_{\mathbf{k}} = 2\mathrm{i} M\big(\sqrt{|\gamma|},\, \mathrm{sgn}(\gamma)\sqrt{|\gamma|}\big),  \\
J_{2\mathbf{k}}^\dagger J_{2\mathbf{k}} &= \gamma\, \mathbf{c}_{\mathbf{k}}^\dagger \sigma^y \mathbf{c}_{\mathbf{k}} + |\gamma|\, \mathbf{c}_{\mathbf{k}}^\dagger \mathbf{c}_{\mathbf{k}} = 2\mathrm{i} M\big(\sqrt{|\gamma|},\, -\mathrm{i}\,\mathrm{sgn}(\gamma)\sqrt{|\gamma|}\big), \\
J_{3\mathbf{k}}^\dagger J_{3\mathbf{k}} &=
\begin{cases}
\gamma\, \mathbf{c}_{\mathbf{k}}^\dagger \sigma^z \mathbf{c}_{\mathbf{k}} + \gamma\, \mathbf{c}_{\mathbf{k}}^\dagger \mathbf{c}_{\mathbf{k}} = 2\mathrm{i} M\big(\sqrt{2|\gamma|},\, 0\big) & : \gamma > 0, \\
\gamma\, \mathbf{c}_{\mathbf{k}}^\dagger \sigma^z \mathbf{c}_{\mathbf{k}} - \gamma\, \mathbf{c}_{\mathbf{k}}^\dagger \mathbf{c}_{\mathbf{k}} = 2\mathrm{i} M\big(0,\, \sqrt{2|\gamma|}\big) & : \gamma < 0.
\end{cases}
\end{align}

We use this decomposition to find an open system, which gives rise to the Rice-Mele Hamiltonian. An open system, whose Hamiltonian is that of the non-Hermitian Rice-Mele model, has only one jump operator given by $\mathbf{a}_{1\textbf{k}} = \big(\sqrt{\gamma}, \text{i}\sqrt{\gamma}\big)$.
The resulting effective non-Hermitian Hamiltonian is then given by Eq.~\eqref{eq:NH-RM}, and the Keldysh self-energy reads~\cite{Talkington:2024gf}
\begin{equation}
    \Sigma^K = \text{i}\gamma\sigma^y = \begin{pmatrix}
        0 & \gamma \\
        -\gamma & 0
    \end{pmatrix}.
\end{equation}

We now proceed to evaluate the quantum response of the introduced open system. Formally, a paramagnetic response polarization bubble (see Fig.~\ref{fig:bubble}) of an open quantum system governed by Lindbladian $\mathcal{L}$ amounts to~\cite{Talkington:2024gf}
\begin{equation}\label{eq::pi}
\begin{aligned}
    \pi_{O_i O_j}(\om) = &-\frac{\text{i}}{2V} \sum_{n,m} \big[h^+_{nm}(\om) + h^-_{mn}(\om) \big] O^i_{nm} O^j_{mn},
\end{aligned}
\text{with}
\end{equation}
\beq{eq:h}
    h^\pm_{nm}(\om)\equiv \int \dd \om' G^{R(A)}_m(\om'\pm \om) G^K_n(\om').
\eeq
}
{${G^{R}_n(\om) = \text{1}/(\om-\varepsilon_n)}$ and ${G^{A}_n(\om) = \text{1}/(\om-\varepsilon^*_n)}$ are \mbox{retarded} and advanced Green's functions,} with ${\varepsilon_n \equiv E_n - \text{i} \Sigma_n''}$ the complex energy eigenvalue of the NH Hamiltonian ${H}$, and ${G^K_m(\om) =G_m^R(\om) \Sigma_m^K(\omega) G_m^A(\om)}$ are Keldysh Green's functions, projected onto states $n,m$, respectively. Above, we assume that the Green's function matrices commute; ${[{H}^{(\dagger)},\Sigma^K] =0}$, and drop the momentum \mbox{parameter} index $\textbf{k}$ for brevity. 

{\color{black} Importantly, we find that under these conditions and for $J_{\lambda\vec{k}} = \vec{a}_{\lambda\vec{k}}\cdot\vec{c}_\vec{k}$, Eq.~\eqref{eq::pi} reduces to a Lorentzian $L_{nm}(\omega)$. {\color{black} Namely, in this case, we have $\Sigma^K=2M$~\cite{Talkington:2024gf}. As such, the~computation of the response is simplified with
\begin{equation}
    G^K(\om) = \frac{1}{\om^2}\frac{1}{1-\om^{-1}{H}}({H}-{H}^\dagger)\frac{1}{1-\om^{-1}{H}^\dagger}.
\end{equation}
Then, upon expanding the Green's functions in power series in $\om^{-1}{H}$, we find
\begin{equation}\label{eq:GK}
    G^K({\om}) = \frac{1}{\om}\lb\frac{{H}}{\om - {H}} - \frac{{H}^\dagger}{\om-{H}^\dagger}\rb = G^R({\om}) - G^A({\om}).
\end{equation}
We substitute Eq.~\eqref{eq:GK} into the expression for response, Eq.~\eqref{eq::pi}. From the pole structure of the obtained terms, we find that one integral in Eq.~\eqref{eq::pi} vanishes identically, as it has no poles in the upper half-plane. We compute the leftover integral simply by entering an identity of the form $\mathbbm{1} = \sum_n\ket{\psi^L_n}\bra{\psi^R_n}$ to evaluate the trace in Eq.~\eqref{eq::pi} as~$\frac{1}{2}\sum_{nml}O^i_{nm}G^R_{ml}(\varepsilon_n^*-\om)O^j_{ln}$.}

 Upon substituting the definition of the retarded Green's function matrix elements $G^R_{ml}(\varepsilon_n^*-\om)$ and evaluating the absorptive part, $\pi_{ij}^\text{abs}(\om) = \frac{1}{2\text{i}}[ \pi_{ij}(\om) - \pi_{ji}(\om)^* ]$, Eq.~\eqref{eq::pi} \mbox{reproduces} the bound Eq.~\eqref{eq:posdef}, as we algebraically detail in~the~SM~\cite{Supplemental}.}

Centrally to this work, we find that for the quantum geometric bounds to be preserved, it is sufficient that,
{\color{black}
\beq{eq:hBound}
    h_{nm}(\om) \equiv -\frac{1}{2} \text{Re} \Big( h^+_{nm}(\om) +  h^-_{mn}(\om) \Big) \geqslant 0 \qquad (E_n < E_m),
\eeq
}
for any considered frequency $\om$. {As a sufficiency condition, Eq.~\eqref{eq:hBound} follows from (i) the positive-semidefiniteness of the $O^i_{nm} O^j_{mn}$ products associated with the generalized Hermitian QGTs~\cite{Shinada2025}, and (ii) the diagonality of the Keldysh Green's function $G^K_m(\om)$ \mbox{under} the assumed commutation relations. Accordingly, for any vector $\textbf{v}$ with complex coefficients $v_i$, ${v^*_i \pi^{\text{abs}}_{O_i O_j}  v_j = \sum_{n,m} h_{nm} O^i_{nm} O^j_{mn} v^*_i v_j = \sum_{n,m} h_{nm} |O^j_{mn}v_j|^2 \geqslant 0}$ {for $\om > 0$, upon pairing $(m,n)$ and $(n,m)$ terms and using $L_{nm}(\om) > L_{mn}(\om)$}~\cite{Supplemental}, with the absorptive part  $\pi^{\text{abs}}_{O_i O_j} \equiv \frac{1}{2 \text{i}} [\pi_{O_i O_j} - (\pi_{O_j O_i})^*]$.} In~\mbox{particular}, these conditions are naturally realized in the NH RM model dynamics and {bounds with dissipation parameters $\gamma, \Gamma$ [see Fig.~\ref{fig:NH}(c)].}

\sect{Discussion and conclusion} We now further discuss the range and applicability of our findings. Importantly, we stress that our bounds are adaptable vastly beyond the pseudo-Hermitian paradigms of~$\mathcal{PT}$ symmetric systems with real eigenvalue spectra~\cite{Zhu2021}. We {stress} that although we demonstrated the NH bound in a~two-band system, our findings are directly generalizable to $N$-band NH systems, upon imposing further constraints on all NH spectral eigenvalues. {In particular, the $N$-band generalization applies when all NH spectral eigenvalue differences $z_m \equiv \varepsilon_{m \kv} - \varepsilon_{0 \kv}$ (${m = 1,\ldots,N-1}$) belong to the lower half-plane of the complex plane, i.e., there exists a band $n=0$ with strictly the lowest decay rate given by ${\Sigma''_0 < \Sigma''_m}$. The $N$-band non-Hermitian Hamiltonian $H$ must further satisfy a~commutation relation $[ H, H^\dagger ] = 0$, i.e., the self-energy matrix $\Sigma = \text{diag}({\Sigma_0, \ldots, \Sigma_{N-1}})$ is band-diagonalizable, analogously to how the Keldysh Green's function $G^K_m(\omega)$ must be band-diagonalizable to obtain the geometric bounds for the \mbox{responses} of the open quantum systems. For further details on the demonstration of the $N$-band generalized non-Hermitian quantum geometric bounds, see the SM~\cite{Supplemental}.} In~fact, the~breakdown of the NH bounds in general multiband NH systems is intuitively expected {when there is no unique slowest-decaying band. Physically, this is \mbox{because} a relative gain of a single band eigenvalue, even in part of the momentum space, can compensate for the combined decays or relative losses due to the other bands in the NH ground state.} Such compensation can culminate in a~cancellation of constituent NH responses, thus trivializing the NH geometric bounds. Remarkably, our general bounds for NH topological systems, can be directly \mbox{validated} in photonic or optomechanical setups simulating NH Hamiltonians~\cite{delPino2022}.

From the perspective of open quantum systems, we stress that the breakdown of geometric bounds can naturally arise upon introducing population inversions in the bath energy level distributions, which could impose an inverse condition, ${h_{nm}<0}$. As long as the statistics of the bath remains uninverted, the positivity of the distribution functions underpinned by the bubble diagrams is naturally ensured, consistently with the causal character of the responses preserved in the Lindbladian formalism. Combined with the projective Hilbert-space structure of the Hermitian QGTs, as encoded in the quantum operators ($O^i_{nm} O^j_{mn}$), the positive distribution functions that impose $h_{nm}\geq0$ further preserve the positive semidefiniteness conditions, and the geometric bounds hold. As a special case, we note that any thermal bath maintains the positivity of geometric bounds, consistent with the setups of experimentally probed solid materials.\\

In this Letter, we retrieved general observable quantum geometric bounds in response functions of non-Hermitian and open quantum systems.
Our findings apply to experimental systems that are intrinsically open quantum systems, and thus, should be governed by the geometric conditions elucidated in this work.
By combining the Keldysh formalism with quantum geometric tensors, this work could initiate a plethora of further studies aimed at identifying the role of open quantum system geometries beyond metric tensors.
Future works can address these general geometric aspects of open quantum systems, with relevance from condensed matter systems to quantum information contexts.

\sect{Note Added} Recently, we became aware of Ref.~\cite{Montag2025} demonstrating lower bounds on the localization of non-Hermitian Wannier functions due to the non-Hermitian quantum metric.
\newline

\begin{acknowledgments}
\sect{Acknowledgments} J.B. thanks Roni Ilan and Moshe Goldstein for insightful discussions. M.M.~\mbox{acknowledges} funding from the James Walters Studentship at \mbox{Trinity College,} Cambridge, and financial support by the Deutsche Forschungsgemeinschaft (DFG, German \mbox{Research Foundation}) via the project A07 of the \mbox{Collaborative} \mbox{Research Center} SFB 1143 (Project No.~247310070). W.J.J. acknowledges funding from the Rod Smallwood Studentship at Trinity College, Cambridge.
J.B.~was supported by EPSRC Grant No.\ EP/V062654/1, a~Leverhulme Early Career Fellowship and the Newton Trust of the University of Cambridge.
\end{acknowledgments}

\bibliography{references}

\begin{thebibliography}{10}%
\makeatletter
\providecommand \@ifxundefined [1]{%
 \@ifx{#1\undefined}
}%
\providecommand \@ifnum [1]{%
 \ifnum #1\expandafter \@firstoftwo
 \else \expandafter \@secondoftwo
 \fi
}%
\providecommand \@ifx [1]{%
 \ifx #1\expandafter \@firstoftwo
 \else \expandafter \@secondoftwo
 \fi
}%
\providecommand \natexlab [1]{#1}%
\providecommand \enquote  [1]{``#1''}%
\providecommand \bibnamefont  [1]{#1}%
\providecommand \bibfnamefont [1]{#1}%
\providecommand \citenamefont [1]{#1}%
\providecommand \href@noop [0]{\@secondoftwo}%
\providecommand \href [0]{\begingroup \@sanitize@url \@href}%
\providecommand \@href[1]{\@@startlink{#1}\@@href}%
\providecommand \@@href[1]{\endgroup#1\@@endlink}%
\providecommand \@sanitize@url [0]{\catcode `\\12\catcode `\$12\catcode `\&12\catcode `\#12\catcode `\^12\catcode `\_12\catcode `\%12\relax}%
\providecommand \@@startlink[1]{}%
\providecommand \@@endlink[0]{}%
\providecommand \url  [0]{\begingroup\@sanitize@url \@url }%
\providecommand \@url [1]{\endgroup\@href {#1}{\urlprefix }}%
\providecommand \urlprefix  [0]{URL }%
\providecommand \Eprint [0]{\href }%
\providecommand \doibase [0]{http://dx.doi.org/}%
\providecommand \selectlanguage [0]{\@gobble}%
\providecommand \bibinfo  [0]{\@secondoftwo}%
\providecommand \bibfield  [0]{\@secondoftwo}%
\providecommand \translation [1]{[#1]}%
\providecommand \BibitemOpen [0]{}%
\providecommand \bibitemStop [0]{}%
\providecommand \bibitemNoStop [0]{.\EOS\space}%
\providecommand \EOS [0]{\spacefactor3000\relax}%
\providecommand \BibitemShut  [1]{\csname bibitem#1\endcsname}%
\let\auto@bib@innerbib\@empty
\bibitem [{\citenamefont {T\"orm\"a}(2023)}]{Torma2023}%
  \BibitemOpen
  \bibfield  {author} {\bibinfo {author} {\bibfnamefont {P.}~\bibnamefont {T\"orm\"a}},\ }\href {https://link.aps.org/doi/10.1103/PhysRevLett.131.240001} {\bibfield  {journal} {\bibinfo  {journal} {Phys. Rev. Lett.}\ }\textbf {\bibinfo {volume} {131}},\ \bibinfo {pages} {240001} (\bibinfo {year} {2023})}\BibitemShut {NoStop}%
\bibitem [{\citenamefont {Silberstein}\ \emph {et~al.}(2020)\citenamefont {Silberstein}, \citenamefont {Behrends}, \citenamefont {Goldstein},\ and\ \citenamefont {Ilan}}]{Silberstein:2020hi}%
  \BibitemOpen
  \bibfield  {author} {\bibinfo {author} {\bibfnamefont {N.}~\bibnamefont {Silberstein}}, \bibinfo {author} {\bibfnamefont {J.}~\bibnamefont {Behrends}}, \bibinfo {author} {\bibfnamefont {M.}~\bibnamefont {Goldstein}}, \ and\ \bibinfo {author} {\bibfnamefont {R.}~\bibnamefont {Ilan}},\ }\href {\doibase 10.1103/PhysRevB.102.245147} {\bibfield  {journal} {\bibinfo  {journal} {Phys. Rev. B}\ }\textbf {\bibinfo {volume} {102}},\ \bibinfo {pages} {245147} (\bibinfo {year} {2020})}\BibitemShut {NoStop}%
\bibitem [{\citenamefont {Behrends}\ \emph {et~al.}()\citenamefont {Behrends}, \citenamefont {Ilan},\ and\ \citenamefont {Goldstein}}]{Behrends2025}%
  \BibitemOpen
  \bibfield  {author} {\bibinfo {author} {\bibfnamefont {J.}~\bibnamefont {Behrends}}, \bibinfo {author} {\bibfnamefont {R.}~\bibnamefont {Ilan}}, \ and\ \bibinfo {author} {\bibfnamefont {M.}~\bibnamefont {Goldstein}},\ }\href@noop {} {\enquote {\bibinfo {title} {Quantum geometry of non-{H}ermitian systems},}\ }\Eprint {http://arxiv.org/abs/2503.13604} {arXiv:2503.13604} \BibitemShut {NoStop}%
\bibitem [{\citenamefont {Onishi}\ and\ \citenamefont {Fu}(2024)}]{Onishi:2024gl}%
  \BibitemOpen
  \bibfield  {author} {\bibinfo {author} {\bibfnamefont {Y.}~\bibnamefont {Onishi}}\ and\ \bibinfo {author} {\bibfnamefont {L.}~\bibnamefont {Fu}},\ }\href {\doibase 10.1103/PhysRevX.14.011052} {\bibfield  {journal} {\bibinfo  {journal} {Phys. Rev. X}\ }\textbf {\bibinfo {volume} {14}},\ \bibinfo {pages} {011052} (\bibinfo {year} {2024})}\BibitemShut {NoStop}%
\bibitem [{\citenamefont {Kudinov}(1991)}]{Kudinov1991}%
  \BibitemOpen
  \bibfield  {author} {\bibinfo {author} {\bibfnamefont {E.~K.}\ \bibnamefont {Kudinov}},\ }\href@noop {} {\bibfield  {journal} {\bibinfo  {journal} {Fiz. Tverd. Tela}\ }\textbf {\bibinfo {volume} {33}},\ \bibinfo {pages} {2306} (\bibinfo {year} {1991})},\ \bibinfo {note} {[Sov.\ Phys.\ Solid State 33, 1299 (1991)]}\BibitemShut {NoStop}%
\bibitem [{\citenamefont {Souza}\ \emph {et~al.}(2000)\citenamefont {Souza}, \citenamefont {Wilkens},\ and\ \citenamefont {Martin}}]{Souza:2000cj}%
  \BibitemOpen
  \bibfield  {author} {\bibinfo {author} {\bibfnamefont {I.}~\bibnamefont {Souza}}, \bibinfo {author} {\bibfnamefont {T.}~\bibnamefont {Wilkens}}, \ and\ \bibinfo {author} {\bibfnamefont {R.~M.}\ \bibnamefont {Martin}},\ }\href {\doibase 10.1103/PhysRevB.62.1666} {\bibfield  {journal} {\bibinfo  {journal} {Phys. Rev. B}\ }\textbf {\bibinfo {volume} {62}},\ \bibinfo {pages} {1666} (\bibinfo {year} {2000})}\BibitemShut {NoStop}%
\bibitem [{\citenamefont {Shen}\ \emph {et~al.}(2018)\citenamefont {Shen}, \citenamefont {Zhen},\ and\ \citenamefont {Fu}}]{Shen2018}%
  \BibitemOpen
  \bibfield  {author} {\bibinfo {author} {\bibfnamefont {H.}~\bibnamefont {Shen}}, \bibinfo {author} {\bibfnamefont {B.}~\bibnamefont {Zhen}}, \ and\ \bibinfo {author} {\bibfnamefont {L.}~\bibnamefont {Fu}},\ }\href {\doibase 10.1103/PhysRevLett.120.146402} {\bibfield  {journal} {\bibinfo  {journal} {Phys. Rev. Lett.}\ }\textbf {\bibinfo {volume} {120}},\ \bibinfo {pages} {146402} (\bibinfo {year} {2018})}\BibitemShut {NoStop}%
\bibitem [{\citenamefont {Daley}(2014)}]{Daley:2014hc}%
  \BibitemOpen
  \bibfield  {author} {\bibinfo {author} {\bibfnamefont {A.~J.}\ \bibnamefont {Daley}},\ }\href {\doibase 10.1080/00018732.2014.933502} {\bibfield  {journal} {\bibinfo  {journal} {Advances in Physics}\ }\textbf {\bibinfo {volume} {63}},\ \bibinfo {pages} {77} (\bibinfo {year} {2014})}\BibitemShut {NoStop}%
\bibitem [{\citenamefont {Monkman}\ and\ \citenamefont {Berciu}(2026)}]{Monkman2025}%
  \BibitemOpen
  \bibfield  {author} {\bibinfo {author} {\bibfnamefont {K.}~\bibnamefont {Monkman}}\ and\ \bibinfo {author} {\bibfnamefont {M.}~\bibnamefont {Berciu}},\ }\href {\doibase 10.1103/dyjw-1cbl} {\bibfield  {journal} {\bibinfo  {journal} {Phys. Rev. A}\ }\textbf {\bibinfo {volume} {113}},\ \bibinfo {pages} {032213} (\bibinfo {year} {2026})}\BibitemShut {NoStop}%
\bibitem [{\citenamefont {Talkington}\ and\ \citenamefont {Claassen}(2024)}]{Talkington:2024gf}%
  \BibitemOpen
  \bibfield  {author} {\bibinfo {author} {\bibfnamefont {S.}~\bibnamefont {Talkington}}\ and\ \bibinfo {author} {\bibfnamefont {M.}~\bibnamefont {Claassen}},\ }\href {\doibase 10.1038/s41535-024-00709-4} {\bibfield  {journal} {\bibinfo  {journal} {npj Quantum Mater.}\ }\textbf {\bibinfo {volume} {9}},\ \bibinfo {pages} {104} (\bibinfo {year} {2024})}\BibitemShut {NoStop}%
\end{thebibliography}%


\begin{thebibliography}{78}%
\makeatletter
\providecommand \@ifxundefined [1]{%
 \@ifx{#1\undefined}
}%
\providecommand \@ifnum [1]{%
 \ifnum #1\expandafter \@firstoftwo
 \else \expandafter \@secondoftwo
 \fi
}%
\providecommand \@ifx [1]{%
 \ifx #1\expandafter \@firstoftwo
 \else \expandafter \@secondoftwo
 \fi
}%
\providecommand \natexlab [1]{#1}%
\providecommand \enquote  [1]{``#1''}%
\providecommand \bibnamefont  [1]{#1}%
\providecommand \bibfnamefont [1]{#1}%
\providecommand \citenamefont [1]{#1}%
\providecommand \href@noop [0]{\@secondoftwo}%
\providecommand \href [0]{\begingroup \@sanitize@url \@href}%
\providecommand \@href[1]{\@@startlink{#1}\@@href}%
\providecommand \@@href[1]{\endgroup#1\@@endlink}%
\providecommand \@sanitize@url [0]{\catcode `\\12\catcode `\$12\catcode `\&12\catcode `\#12\catcode `\^12\catcode `\_12\catcode `\%12\relax}%
\providecommand \@@startlink[1]{}%
\providecommand \@@endlink[0]{}%
\providecommand \url  [0]{\begingroup\@sanitize@url \@url }%
\providecommand \@url [1]{\endgroup\@href {#1}{\urlprefix }}%
\providecommand \urlprefix  [0]{URL }%
\providecommand \Eprint [0]{\href }%
\providecommand \doibase [0]{https://doi.org/}%
\providecommand \selectlanguage [0]{\@gobble}%
\providecommand \bibinfo  [0]{\@secondoftwo}%
\providecommand \bibfield  [0]{\@secondoftwo}%
\providecommand \translation [1]{[#1]}%
\providecommand \BibitemOpen [0]{}%
\providecommand \bibitemStop [0]{}%
\providecommand \bibitemNoStop [0]{.\EOS\space}%
\providecommand \EOS [0]{\spacefactor3000\relax}%
\providecommand \BibitemShut  [1]{\csname bibitem#1\endcsname}%
\let\auto@bib@innerbib\@empty
\bibitem [{\citenamefont {Libermann}\ and\ \citenamefont {Marle}(1987)}]{analytical_mechanics}%
  \BibitemOpen
  \bibfield  {author} {\bibinfo {author} {\bibfnamefont {P.}~\bibnamefont {Libermann}}\ and\ \bibinfo {author} {\bibfnamefont {C.-M.}\ \bibnamefont {Marle}},\ }\href {https://doi.org/10.1007/978-94-009-3807-6} {\emph {\bibinfo {title} {{Symplectic Geometry and Analytical Mechanics}}}}\ (\bibinfo  {publisher} {Springer Dordrecht},\ \bibinfo {address} {Dordrecht, Holland},\ \bibinfo {year} {1987})\BibitemShut {NoStop}%
\bibitem [{\citenamefont {O'Neill}(1983)}]{o1983semi}%
  \BibitemOpen
  \bibfield  {author} {\bibinfo {author} {\bibfnamefont {B.}~\bibnamefont {O'Neill}},\ }\href@noop {} {\emph {\bibinfo {title} {Semi-Riemannian Geometry With Applications to Relativity}}},\ Pure and Applied Mathematics\ (\bibinfo  {publisher} {Academic Press},\ \bibinfo {address} {Cambridge, M.A.},\ \bibinfo {year} {1983})\BibitemShut {NoStop}%
\bibitem [{\citenamefont {Provost}\ and\ \citenamefont {Vallee}(1980)}]{Provost1980}%
  \BibitemOpen
  \bibfield  {author} {\bibinfo {author} {\bibfnamefont {J.~P.}\ \bibnamefont {Provost}}\ and\ \bibinfo {author} {\bibfnamefont {G.}~\bibnamefont {Vallee}},\ }\bibfield  {title} {\bibinfo {title} {Riemannian structure on manifolds of quantum states},\ }\href {https://doi.org/10.1007/BF02193559} {\bibfield  {journal} {\bibinfo  {journal} {Commun. Math. Phys.}\ }\textbf {\bibinfo {volume} {76}},\ \bibinfo {pages} {289} (\bibinfo {year} {1980})}\BibitemShut {NoStop}%
\bibitem [{\citenamefont {Anandan}\ and\ \citenamefont {Aharonov}(1990)}]{Anandan:1990dx}%
  \BibitemOpen
  \bibfield  {author} {\bibinfo {author} {\bibfnamefont {J.}~\bibnamefont {Anandan}}\ and\ \bibinfo {author} {\bibfnamefont {Y.}~\bibnamefont {Aharonov}},\ }\bibfield  {title} {\bibinfo {title} {{Geometry of quantum evolution}},\ }\href {https://doi.org/10.1103/PhysRevLett.65.1697} {\bibfield  {journal} {\bibinfo  {journal} {Phys. Rev. Lett.}\ }\textbf {\bibinfo {volume} {65}},\ \bibinfo {pages} {1697} (\bibinfo {year} {1990})}\BibitemShut {NoStop}%
\bibitem [{\citenamefont {Kolodrubetz}\ \emph {et~al.}(2017)\citenamefont {Kolodrubetz}, \citenamefont {Sels}, \citenamefont {Mehta},\ and\ \citenamefont {Polkovnikov}}]{Kolodrubetz:2017jg}%
  \BibitemOpen
  \bibfield  {author} {\bibinfo {author} {\bibfnamefont {M.}~\bibnamefont {Kolodrubetz}}, \bibinfo {author} {\bibfnamefont {D.}~\bibnamefont {Sels}}, \bibinfo {author} {\bibfnamefont {P.}~\bibnamefont {Mehta}},\ and\ \bibinfo {author} {\bibfnamefont {A.}~\bibnamefont {Polkovnikov}},\ }\bibfield  {title} {\bibinfo {title} {{Geometry and non-adiabatic response in quantum and classical systems}},\ }\href {https://doi.org/10.1016/j.physrep.2017.07.001} {\bibfield  {journal} {\bibinfo  {journal} {Phys. Rep.}\ }\textbf {\bibinfo {volume} {697}},\ \bibinfo {pages} {1} (\bibinfo {year} {2017})}\BibitemShut {NoStop}%
\bibitem [{\citenamefont {T\"orm\"a}(2023)}]{Torma2023}%
  \BibitemOpen
  \bibfield  {author} {\bibinfo {author} {\bibfnamefont {P.}~\bibnamefont {T\"orm\"a}},\ }\bibfield  {title} {\bibinfo {title} {{Essay: Where Can Quantum Geometry Lead Us?}},\ }\href {https://link.aps.org/doi/10.1103/PhysRevLett.131.240001} {\bibfield  {journal} {\bibinfo  {journal} {Phys. Rev. Lett.}\ }\textbf {\bibinfo {volume} {131}},\ \bibinfo {pages} {240001} (\bibinfo {year} {2023})}\BibitemShut {NoStop}%
\bibitem [{\citenamefont {Verma}\ \emph {et~al.}(2026)\citenamefont {Verma}, \citenamefont {Moll}, \citenamefont {Holder},\ and\ \citenamefont {Queiroz}}]{Verma2025}%
  \BibitemOpen
  \bibfield  {author} {\bibinfo {author} {\bibfnamefont {N.}~\bibnamefont {Verma}}, \bibinfo {author} {\bibfnamefont {P.~J.~W.}\ \bibnamefont {Moll}}, \bibinfo {author} {\bibfnamefont {T.}~\bibnamefont {Holder}},\ and\ \bibinfo {author} {\bibfnamefont {R.}~\bibnamefont {Queiroz}},\ }\bibfield  {title} {\bibinfo {title} {Quantum geometry and the hidden scales in materials},\ }\href {https://doi.org/10.1038/s42254-026-00923-y} {\bibfield  {journal} {\bibinfo  {journal} {Nat. Rev. Phys.}\ }\textbf {\bibinfo {volume} {8}},\ \bibinfo {pages} {226–239} (\bibinfo {year} {2026})}\BibitemShut {NoStop}%
\bibitem [{\citenamefont {Facchi}\ \emph {et~al.}(2010)\citenamefont {Facchi}, \citenamefont {Kulkarni}, \citenamefont {Man'ko}, \citenamefont {Marmo}, \citenamefont {Sudarshan},\ and\ \citenamefont {Ventriglia}}]{Facchi:2010hl}%
  \BibitemOpen
  \bibfield  {author} {\bibinfo {author} {\bibfnamefont {P.}~\bibnamefont {Facchi}}, \bibinfo {author} {\bibfnamefont {R.}~\bibnamefont {Kulkarni}}, \bibinfo {author} {\bibfnamefont {V.}~\bibnamefont {Man'ko}}, \bibinfo {author} {\bibfnamefont {G.}~\bibnamefont {Marmo}}, \bibinfo {author} {\bibfnamefont {E.}~\bibnamefont {Sudarshan}},\ and\ \bibinfo {author} {\bibfnamefont {F.}~\bibnamefont {Ventriglia}},\ }\bibfield  {title} {\bibinfo {title} {{Classical and quantum Fisher information in the geometrical formulation of quantum mechanics}},\ }\href {https://doi.org/10.1016/j.physleta.2010.10.005} {\bibfield  {journal} {\bibinfo  {journal} {Phys. Lett. A}\ }\textbf {\bibinfo {volume} {374}},\ \bibinfo {pages} {4801} (\bibinfo {year} {2010})}\BibitemShut {NoStop}%
\bibitem [{\citenamefont {Liu}\ \emph {et~al.}(2019)\citenamefont {Liu}, \citenamefont {Yuan}, \citenamefont {Lu},\ and\ \citenamefont {Wang}}]{Liu2019}%
  \BibitemOpen
  \bibfield  {author} {\bibinfo {author} {\bibfnamefont {J.}~\bibnamefont {Liu}}, \bibinfo {author} {\bibfnamefont {H.}~\bibnamefont {Yuan}}, \bibinfo {author} {\bibfnamefont {X.-M.}\ \bibnamefont {Lu}},\ and\ \bibinfo {author} {\bibfnamefont {X.}~\bibnamefont {Wang}},\ }\bibfield  {title} {\bibinfo {title} {{Quantum Fisher information matrix and multiparameter estimation}},\ }\href {https://dx.doi.org/10.1088/1751-8121/ab5d4d} {\bibfield  {journal} {\bibinfo  {journal} {J. Phys. A: Math. Theor.}\ }\textbf {\bibinfo {volume} {53}},\ \bibinfo {pages} {023001} (\bibinfo {year} {2019})}\BibitemShut {NoStop}%
\bibitem [{\citenamefont {Yu}\ \emph {et~al.}(2022)\citenamefont {Yu}, \citenamefont {Liu}, \citenamefont {Yang}, \citenamefont {Gong}, \citenamefont {Cao}, \citenamefont {Zhang}, \citenamefont {Liu}, \citenamefont {Heyl}, \citenamefont {Ozawa}, \citenamefont {Goldman},\ and\ \citenamefont {Cai}}]{Yu2022}%
  \BibitemOpen
  \bibfield  {author} {\bibinfo {author} {\bibfnamefont {M.}~\bibnamefont {Yu}}, \bibinfo {author} {\bibfnamefont {Y.}~\bibnamefont {Liu}}, \bibinfo {author} {\bibfnamefont {P.}~\bibnamefont {Yang}}, \bibinfo {author} {\bibfnamefont {M.}~\bibnamefont {Gong}}, \bibinfo {author} {\bibfnamefont {Q.}~\bibnamefont {Cao}}, \bibinfo {author} {\bibfnamefont {S.}~\bibnamefont {Zhang}}, \bibinfo {author} {\bibfnamefont {H.}~\bibnamefont {Liu}}, \bibinfo {author} {\bibfnamefont {M.}~\bibnamefont {Heyl}}, \bibinfo {author} {\bibfnamefont {T.}~\bibnamefont {Ozawa}}, \bibinfo {author} {\bibfnamefont {N.}~\bibnamefont {Goldman}},\ and\ \bibinfo {author} {\bibfnamefont {J.}~\bibnamefont {Cai}},\ }\bibfield  {title} {\bibinfo {title} {Quantum {F}isher information measurement and verification of the quantum {C}ram{\'e}r--{R}ao bound in a solid-state qubit},\ }\href {https://doi.org/10.1038/s41534-022-00547-x} {\bibfield  {journal} {\bibinfo  {journal} {npj Quantum Inf.}\ }\textbf {\bibinfo {volume} {8}},\ \bibinfo {pages} {56}
  (\bibinfo {year} {2022})}\BibitemShut {NoStop}%
\bibitem [{\citenamefont {Bouhon}\ \emph {et~al.}()\citenamefont {Bouhon}, \citenamefont {Timmel},\ and\ \citenamefont {Slager}}]{Bouhon2023}%
  \BibitemOpen
  \bibfield  {author} {\bibinfo {author} {\bibfnamefont {A.}~\bibnamefont {Bouhon}}, \bibinfo {author} {\bibfnamefont {A.}~\bibnamefont {Timmel}},\ and\ \bibinfo {author} {\bibfnamefont {R.-J.}\ \bibnamefont {Slager}},\ }\href@noop {} {\bibinfo {title} {Quantum geometry beyond projective single bands}},\ \Eprint {https://arxiv.org/abs/2303.02180} {arXiv:2303.02180} \BibitemShut {NoStop}%
\bibitem [{\citenamefont {Yu}\ \emph {et~al.}(2024)\citenamefont {Yu}, \citenamefont {Li}, \citenamefont {Chu}, \citenamefont {Mera}, \citenamefont {{\"U}nal}, \citenamefont {Yang}, \citenamefont {Liu}, \citenamefont {Goldman},\ and\ \citenamefont {Cai}}]{Yu2024a}%
  \BibitemOpen
  \bibfield  {author} {\bibinfo {author} {\bibfnamefont {M.}~\bibnamefont {Yu}}, \bibinfo {author} {\bibfnamefont {X.}~\bibnamefont {Li}}, \bibinfo {author} {\bibfnamefont {Y.}~\bibnamefont {Chu}}, \bibinfo {author} {\bibfnamefont {B.}~\bibnamefont {Mera}}, \bibinfo {author} {\bibfnamefont {F.~N.}\ \bibnamefont {{\"U}nal}}, \bibinfo {author} {\bibfnamefont {P.}~\bibnamefont {Yang}}, \bibinfo {author} {\bibfnamefont {Y.}~\bibnamefont {Liu}}, \bibinfo {author} {\bibfnamefont {N.}~\bibnamefont {Goldman}},\ and\ \bibinfo {author} {\bibfnamefont {J.}~\bibnamefont {Cai}},\ }\bibfield  {title} {\bibinfo {title} {Experimental demonstration of topological bounds in quantum metrology},\ }\href {https://doi.org/10.1093/nsr/nwae065} {\bibfield  {journal} {\bibinfo  {journal} {Nat. Sci. Rev.}\ }\textbf {\bibinfo {volume} {11}},\ \bibinfo {pages} {nwae065} (\bibinfo {year} {2024})}\BibitemShut {NoStop}%
\bibitem [{\citenamefont {Wahl}\ \emph {et~al.}(2025)\citenamefont {Wahl}, \citenamefont {Jankowski}, \citenamefont {Bouhon}, \citenamefont {Chaudhary},\ and\ \citenamefont {Slager}}]{Wahl2025}%
  \BibitemOpen
  \bibfield  {author} {\bibinfo {author} {\bibfnamefont {T.~B.}\ \bibnamefont {Wahl}}, \bibinfo {author} {\bibfnamefont {W.~J.}\ \bibnamefont {Jankowski}}, \bibinfo {author} {\bibfnamefont {A.}~\bibnamefont {Bouhon}}, \bibinfo {author} {\bibfnamefont {G.}~\bibnamefont {Chaudhary}},\ and\ \bibinfo {author} {\bibfnamefont {R.-J.}\ \bibnamefont {Slager}},\ }\bibfield  {title} {\bibinfo {title} {Exact projected entangled pair ground states with topological {E}uler invariant},\ }\href {https://doi.org/10.1038/s41467-024-55484-4} {\bibfield  {journal} {\bibinfo  {journal} {Nat. Commun.}\ }\textbf {\bibinfo {volume} {16}},\ \bibinfo {pages} {284} (\bibinfo {year} {2025})}\BibitemShut {NoStop}%
\bibitem [{\citenamefont {Jankowski}\ \emph {et~al.}(2025{\natexlab{a}})\citenamefont {Jankowski}, \citenamefont {Slager},\ and\ \citenamefont {Lange}}]{Jankowski2025PRR}%
  \BibitemOpen
  \bibfield  {author} {\bibinfo {author} {\bibfnamefont {W.~J.}\ \bibnamefont {Jankowski}}, \bibinfo {author} {\bibfnamefont {R.-J.}\ \bibnamefont {Slager}},\ and\ \bibinfo {author} {\bibfnamefont {G.~F.}\ \bibnamefont {Lange}},\ }\bibfield  {title} {\bibinfo {title} {Quantum geometric bounds in spinful systems with trivial band topology},\ }\href {https://doi.org/10.1103/zlxq-fxgc} {\bibfield  {journal} {\bibinfo  {journal} {Phys. Rev. Res.}\ }\textbf {\bibinfo {volume} {7}},\ \bibinfo {pages} {L042011} (\bibinfo {year} {2025}{\natexlab{a}})}\BibitemShut {NoStop}%
\bibitem [{\citenamefont {Marzari}\ and\ \citenamefont {Vanderbilt}(1997)}]{Marzari1997}%
  \BibitemOpen
  \bibfield  {author} {\bibinfo {author} {\bibfnamefont {N.}~\bibnamefont {Marzari}}\ and\ \bibinfo {author} {\bibfnamefont {D.}~\bibnamefont {Vanderbilt}},\ }\bibfield  {title} {\bibinfo {title} {Maximally localized generalized {W}annier functions for composite energy bands},\ }\href {https://link.aps.org/doi/10.1103/PhysRevB.56.12847} {\bibfield  {journal} {\bibinfo  {journal} {Phys. Rev. B}\ }\textbf {\bibinfo {volume} {56}},\ \bibinfo {pages} {12847} (\bibinfo {year} {1997})}\BibitemShut {NoStop}%
\bibitem [{\citenamefont {Rhim}\ \emph {et~al.}(2020)\citenamefont {Rhim}, \citenamefont {Kim},\ and\ \citenamefont {Yang}}]{Rhim2020}%
  \BibitemOpen
  \bibfield  {author} {\bibinfo {author} {\bibfnamefont {J.-W.}\ \bibnamefont {Rhim}}, \bibinfo {author} {\bibfnamefont {K.}~\bibnamefont {Kim}},\ and\ \bibinfo {author} {\bibfnamefont {B.-J.}\ \bibnamefont {Yang}},\ }\bibfield  {title} {\bibinfo {title} {{Quantum distance and anomalous Landau levels of flat bands}},\ }\href {https://doi.org/10.1038/s41586-020-2540-1} {\bibfield  {journal} {\bibinfo  {journal} {Nature}\ }\textbf {\bibinfo {volume} {584}},\ \bibinfo {pages} {59} (\bibinfo {year} {2020})}\BibitemShut {NoStop}%
\bibitem [{\citenamefont {Peotta}\ and\ \citenamefont {T{\"o}rm{\"a}}(2015)}]{Peotta2015}%
  \BibitemOpen
  \bibfield  {author} {\bibinfo {author} {\bibfnamefont {S.}~\bibnamefont {Peotta}}\ and\ \bibinfo {author} {\bibfnamefont {P.}~\bibnamefont {T{\"o}rm{\"a}}},\ }\bibfield  {title} {\bibinfo {title} {Superfluidity in topologically nontrivial flat bands},\ }\href {https://doi.org/10.1038/ncomms9944} {\bibfield  {journal} {\bibinfo  {journal} {Nat. Commun.}\ }\textbf {\bibinfo {volume} {6}},\ \bibinfo {pages} {8944} (\bibinfo {year} {2015})}\BibitemShut {NoStop}%
\bibitem [{\citenamefont {Xie}\ \emph {et~al.}(2020)\citenamefont {Xie}, \citenamefont {Song}, \citenamefont {Lian},\ and\ \citenamefont {Bernevig}}]{Xie2020}%
  \BibitemOpen
  \bibfield  {author} {\bibinfo {author} {\bibfnamefont {F.}~\bibnamefont {Xie}}, \bibinfo {author} {\bibfnamefont {Z.}~\bibnamefont {Song}}, \bibinfo {author} {\bibfnamefont {B.}~\bibnamefont {Lian}},\ and\ \bibinfo {author} {\bibfnamefont {B.~A.}\ \bibnamefont {Bernevig}},\ }\bibfield  {title} {\bibinfo {title} {Topology-bounded superfluid weight in twisted bilayer graphene},\ }\href {https://link.aps.org/doi/10.1103/PhysRevLett.124.167002} {\bibfield  {journal} {\bibinfo  {journal} {Phys. Rev. Lett.}\ }\textbf {\bibinfo {volume} {124}},\ \bibinfo {pages} {167002} (\bibinfo {year} {2020})}\BibitemShut {NoStop}%
\bibitem [{\citenamefont {Verma}\ \emph {et~al.}(2021)\citenamefont {Verma}, \citenamefont {Hazra},\ and\ \citenamefont {Randeria}}]{Verma2021}%
  \BibitemOpen
  \bibfield  {author} {\bibinfo {author} {\bibfnamefont {N.}~\bibnamefont {Verma}}, \bibinfo {author} {\bibfnamefont {T.}~\bibnamefont {Hazra}},\ and\ \bibinfo {author} {\bibfnamefont {M.}~\bibnamefont {Randeria}},\ }\bibfield  {title} {\bibinfo {title} {{Optical spectral weight, phase stiffness, and $T_C$ bounds for trivial and topological flat band superconductors}},\ }\href {https://doi.org/10.1073/pnas.2106744118} {\bibfield  {journal} {\bibinfo  {journal} {PNAS}\ }\textbf {\bibinfo {volume} {118}},\ \bibinfo {pages} {e2106744118} (\bibinfo {year} {2021})}\BibitemShut {NoStop}%
\bibitem [{\citenamefont {Peri}\ \emph {et~al.}(2021)\citenamefont {Peri}, \citenamefont {Song}, \citenamefont {Bernevig},\ and\ \citenamefont {Huber}}]{Peri2021a}%
  \BibitemOpen
  \bibfield  {author} {\bibinfo {author} {\bibfnamefont {V.}~\bibnamefont {Peri}}, \bibinfo {author} {\bibfnamefont {Z.-D.}\ \bibnamefont {Song}}, \bibinfo {author} {\bibfnamefont {B.~A.}\ \bibnamefont {Bernevig}},\ and\ \bibinfo {author} {\bibfnamefont {S.~D.}\ \bibnamefont {Huber}},\ }\bibfield  {title} {\bibinfo {title} {Fragile topology and flat-band superconductivity in the strong-coupling regime},\ }\href {https://link.aps.org/doi/10.1103/PhysRevLett.126.027002} {\bibfield  {journal} {\bibinfo  {journal} {Phys. Rev. Lett.}\ }\textbf {\bibinfo {volume} {126}},\ \bibinfo {pages} {027002} (\bibinfo {year} {2021})}\BibitemShut {NoStop}%
\bibitem [{\citenamefont {Herzog-Arbeitman}\ \emph {et~al.}(2022)\citenamefont {Herzog-Arbeitman}, \citenamefont {Peri}, \citenamefont {Schindler}, \citenamefont {Huber},\ and\ \citenamefont {Bernevig}}]{HerzogArbeitman2022}%
  \BibitemOpen
  \bibfield  {author} {\bibinfo {author} {\bibfnamefont {J.}~\bibnamefont {Herzog-Arbeitman}}, \bibinfo {author} {\bibfnamefont {V.}~\bibnamefont {Peri}}, \bibinfo {author} {\bibfnamefont {F.}~\bibnamefont {Schindler}}, \bibinfo {author} {\bibfnamefont {S.~D.}\ \bibnamefont {Huber}},\ and\ \bibinfo {author} {\bibfnamefont {B.~A.}\ \bibnamefont {Bernevig}},\ }\bibfield  {title} {\bibinfo {title} {Superfluid weight bounds from symmetry and quantum geometry in flat bands},\ }\href {https://link.aps.org/doi/10.1103/PhysRevLett.128.087002} {\bibfield  {journal} {\bibinfo  {journal} {Phys. Rev. Lett.}\ }\textbf {\bibinfo {volume} {128}},\ \bibinfo {pages} {087002} (\bibinfo {year} {2022})}\BibitemShut {NoStop}%
\bibitem [{\citenamefont {T{\"{o}}rm{\"{a}}}\ \emph {et~al.}(2022)\citenamefont {T{\"{o}}rm{\"{a}}}, \citenamefont {Peotta},\ and\ \citenamefont {Bernevig}}]{Torma:2022jg}%
  \BibitemOpen
  \bibfield  {author} {\bibinfo {author} {\bibfnamefont {P.}~\bibnamefont {T{\"{o}}rm{\"{a}}}}, \bibinfo {author} {\bibfnamefont {S.}~\bibnamefont {Peotta}},\ and\ \bibinfo {author} {\bibfnamefont {B.~A.}\ \bibnamefont {Bernevig}},\ }\bibfield  {title} {\bibinfo {title} {{Superconductivity, superfluidity and quantum geometry in twisted multilayer systems}},\ }\href {https://doi.org/10.1038/s42254-022-00466-y} {\bibfield  {journal} {\bibinfo  {journal} {Nat. Rev. Phys.}\ }\textbf {\bibinfo {volume} {4}},\ \bibinfo {pages} {528} (\bibinfo {year} {2022})}\BibitemShut {NoStop}%
\bibitem [{\citenamefont {Mao}\ and\ \citenamefont {Chowdhury}(2024)}]{Mao2024}%
  \BibitemOpen
  \bibfield  {author} {\bibinfo {author} {\bibfnamefont {D.}~\bibnamefont {Mao}}\ and\ \bibinfo {author} {\bibfnamefont {D.}~\bibnamefont {Chowdhury}},\ }\bibfield  {title} {\bibinfo {title} {Upper bounds on superconducting and excitonic phase stiffness for interacting isolated narrow bands},\ }\href {https://doi.org/10.1103/PhysRevB.109.024507} {\bibfield  {journal} {\bibinfo  {journal} {Phys. Rev. B}\ }\textbf {\bibinfo {volume} {109}},\ \bibinfo {pages} {024507} (\bibinfo {year} {2024})}\BibitemShut {NoStop}%
\bibitem [{\citenamefont {Hu}\ \emph {et~al.}(2024)\citenamefont {Hu}, \citenamefont {Ostrovskaya},\ and\ \citenamefont {Estrecho}}]{Hu:2024kw}%
  \BibitemOpen
  \bibfield  {author} {\bibinfo {author} {\bibfnamefont {Y.-M.~R.}\ \bibnamefont {Hu}}, \bibinfo {author} {\bibfnamefont {E.~A.}\ \bibnamefont {Ostrovskaya}},\ and\ \bibinfo {author} {\bibfnamefont {E.}~\bibnamefont {Estrecho}},\ }\bibfield  {title} {\bibinfo {title} {{Generalized quantum geometric tensor in a non-Hermitian exciton-polariton system [Invited]}},\ }\href {https://doi.org/10.1364/OME.497010} {\bibfield  {journal} {\bibinfo  {journal} {Opt. Mater. Express}\ }\textbf {\bibinfo {volume} {14}},\ \bibinfo {pages} {664} (\bibinfo {year} {2024})}\BibitemShut {NoStop}%
\bibitem [{\citenamefont {Jankowski}\ \emph {et~al.}(2025{\natexlab{b}})\citenamefont {Jankowski}, \citenamefont {Thompson}, \citenamefont {Monserrat},\ and\ \citenamefont {Slager}}]{Jankowski2025exciton}%
  \BibitemOpen
  \bibfield  {author} {\bibinfo {author} {\bibfnamefont {W.~J.}\ \bibnamefont {Jankowski}}, \bibinfo {author} {\bibfnamefont {J.~J.~P.}\ \bibnamefont {Thompson}}, \bibinfo {author} {\bibfnamefont {B.}~\bibnamefont {Monserrat}},\ and\ \bibinfo {author} {\bibfnamefont {R.-J.}\ \bibnamefont {Slager}},\ }\bibfield  {title} {\bibinfo {title} {Excitonic topology and quantum geometry in organic semiconductors},\ }\href {https://doi.org/10.1038/s41467-025-59257-5} {\bibfield  {journal} {\bibinfo  {journal} {Nat. Commun.}\ }\textbf {\bibinfo {volume} {16}},\ \bibinfo {pages} {4661} (\bibinfo {year} {2025}{\natexlab{b}})}\BibitemShut {NoStop}%
\bibitem [{\citenamefont {Thompson}\ \emph {et~al.}(2025)\citenamefont {Thompson}, \citenamefont {Jankowski}, \citenamefont {Slager},\ and\ \citenamefont {Monserrat}}]{Thompson2024exciton}%
  \BibitemOpen
  \bibfield  {author} {\bibinfo {author} {\bibfnamefont {J.~J.~P.}\ \bibnamefont {Thompson}}, \bibinfo {author} {\bibfnamefont {W.~J.}\ \bibnamefont {Jankowski}}, \bibinfo {author} {\bibfnamefont {R.-J.}\ \bibnamefont {Slager}},\ and\ \bibinfo {author} {\bibfnamefont {B.}~\bibnamefont {Monserrat}},\ }\bibfield  {title} {\bibinfo {title} {Topologically enhanced exciton transport},\ }\href {https://doi.org/10.1038/s41467-025-66276-9} {\bibfield  {journal} {\bibinfo  {journal} {Nat. Commun.}\ }\textbf {\bibinfo {volume} {16}},\ \bibinfo {pages} {11448} (\bibinfo {year} {2025})}\BibitemShut {NoStop}%
\bibitem [{\citenamefont {Ahn}\ \emph {et~al.}(2020)\citenamefont {Ahn}, \citenamefont {Guo},\ and\ \citenamefont {Nagaosa}}]{Ahn2020}%
  \BibitemOpen
  \bibfield  {author} {\bibinfo {author} {\bibfnamefont {J.}~\bibnamefont {Ahn}}, \bibinfo {author} {\bibfnamefont {G.-Y.}\ \bibnamefont {Guo}},\ and\ \bibinfo {author} {\bibfnamefont {N.}~\bibnamefont {Nagaosa}},\ }\bibfield  {title} {\bibinfo {title} {Low-frequency divergence and quantum geometry of the bulk photovoltaic effect in topological semimetals},\ }\href {https://link.aps.org/doi/10.1103/PhysRevX.10.041041} {\bibfield  {journal} {\bibinfo  {journal} {Phys. Rev. X}\ }\textbf {\bibinfo {volume} {10}},\ \bibinfo {pages} {041041} (\bibinfo {year} {2020})}\BibitemShut {NoStop}%
\bibitem [{\citenamefont {Ahn}\ \emph {et~al.}(2021)\citenamefont {Ahn}, \citenamefont {Guo}, \citenamefont {Nagaosa},\ and\ \citenamefont {Vishwanath}}]{Ahn2021}%
  \BibitemOpen
  \bibfield  {author} {\bibinfo {author} {\bibfnamefont {J.}~\bibnamefont {Ahn}}, \bibinfo {author} {\bibfnamefont {G.-Y.}\ \bibnamefont {Guo}}, \bibinfo {author} {\bibfnamefont {N.}~\bibnamefont {Nagaosa}},\ and\ \bibinfo {author} {\bibfnamefont {A.}~\bibnamefont {Vishwanath}},\ }\bibfield  {title} {\bibinfo {title} {Riemannian geometry of resonant optical responses},\ }\href {http://dx.doi.org/10.1038/s41567-021-01465-z} {\bibfield  {journal} {\bibinfo  {journal} {Nat. Phys.}\ }\textbf {\bibinfo {volume} {18}},\ \bibinfo {pages} {290–295} (\bibinfo {year} {2021})}\BibitemShut {NoStop}%
\bibitem [{\citenamefont {Onishi}\ and\ \citenamefont {Fu}(2024)}]{Onishi:2024gl}%
  \BibitemOpen
  \bibfield  {author} {\bibinfo {author} {\bibfnamefont {Y.}~\bibnamefont {Onishi}}\ and\ \bibinfo {author} {\bibfnamefont {L.}~\bibnamefont {Fu}},\ }\bibfield  {title} {\bibinfo {title} {{Fundamental Bound on Topological Gap}},\ }\href {https://doi.org/10.1103/PhysRevX.14.011052} {\bibfield  {journal} {\bibinfo  {journal} {Phys. Rev. X}\ }\textbf {\bibinfo {volume} {14}},\ \bibinfo {pages} {011052} (\bibinfo {year} {2024})}\BibitemShut {NoStop}%
\bibitem [{\citenamefont {Jankowski}\ and\ \citenamefont {Slager}(2024)}]{Jankowski2024PRL}%
  \BibitemOpen
  \bibfield  {author} {\bibinfo {author} {\bibfnamefont {W.~J.}\ \bibnamefont {Jankowski}}\ and\ \bibinfo {author} {\bibfnamefont {R.-J.}\ \bibnamefont {Slager}},\ }\bibfield  {title} {\bibinfo {title} {Quantized integrated shift effect in multigap topological phases},\ }\href {https://doi.org/10.1103/PhysRevLett.133.186601} {\bibfield  {journal} {\bibinfo  {journal} {Phys. Rev. Lett.}\ }\textbf {\bibinfo {volume} {133}},\ \bibinfo {pages} {186601} (\bibinfo {year} {2024})}\BibitemShut {NoStop}%
\bibitem [{\citenamefont {Yu}\ \emph {et~al.}(2025)\citenamefont {Yu}, \citenamefont {Herzog-Arbeitman},\ and\ \citenamefont {Bernevig}}]{Yu2025}%
  \BibitemOpen
  \bibfield  {author} {\bibinfo {author} {\bibfnamefont {J.}~\bibnamefont {Yu}}, \bibinfo {author} {\bibfnamefont {J.}~\bibnamefont {Herzog-Arbeitman}},\ and\ \bibinfo {author} {\bibfnamefont {B.~A.}\ \bibnamefont {Bernevig}},\ }\bibfield  {title} {\bibinfo {title} {Universal {W}ilson loop bound of quantum geometry},\ }\href {https://doi.org/10.1103/mp2c-zzkt} {\bibfield  {journal} {\bibinfo  {journal} {Phys. Rev. Lett.}\ }\textbf {\bibinfo {volume} {135}},\ \bibinfo {pages} {086401} (\bibinfo {year} {2025})}\BibitemShut {NoStop}%
\bibitem [{\citenamefont {Jankowski}\ \emph {et~al.}(2025{\natexlab{c}})\citenamefont {Jankowski}, \citenamefont {Slager},\ and\ \citenamefont {Pizzochero}}]{Jankowski2025PRL}%
  \BibitemOpen
  \bibfield  {author} {\bibinfo {author} {\bibfnamefont {W.~J.}\ \bibnamefont {Jankowski}}, \bibinfo {author} {\bibfnamefont {R.-J.}\ \bibnamefont {Slager}},\ and\ \bibinfo {author} {\bibfnamefont {M.}~\bibnamefont {Pizzochero}},\ }\bibfield  {title} {\bibinfo {title} {Enhancing the hyperpolarizability of crystals with quantum geometry},\ }\href {https://doi.org/10.1103/z7lp-pqp6} {\bibfield  {journal} {\bibinfo  {journal} {Phys. Rev. Lett.}\ }\textbf {\bibinfo {volume} {135}},\ \bibinfo {pages} {126606} (\bibinfo {year} {2025}{\natexlab{c}})}\BibitemShut {NoStop}%
\bibitem [{\citenamefont {Kang}\ \emph {et~al.}(2025)\citenamefont {Kang}, \citenamefont {Kim}, \citenamefont {Qian}, \citenamefont {Neves}, \citenamefont {Ye}, \citenamefont {Jung}, \citenamefont {Puntel}, \citenamefont {Mazzola}, \citenamefont {Fang}, \citenamefont {Jozwiak}, \citenamefont {Bostwick}, \citenamefont {Rotenberg}, \citenamefont {Fuji}, \citenamefont {Vobornik}, \citenamefont {Park}, \citenamefont {Checkelsky}, \citenamefont {Yang},\ and\ \citenamefont {Comin}}]{Kang2025}%
  \BibitemOpen
  \bibfield  {author} {\bibinfo {author} {\bibfnamefont {M.}~\bibnamefont {Kang}}, \bibinfo {author} {\bibfnamefont {S.}~\bibnamefont {Kim}}, \bibinfo {author} {\bibfnamefont {Y.}~\bibnamefont {Qian}}, \bibinfo {author} {\bibfnamefont {P.~M.}\ \bibnamefont {Neves}}, \bibinfo {author} {\bibfnamefont {L.}~\bibnamefont {Ye}}, \bibinfo {author} {\bibfnamefont {J.}~\bibnamefont {Jung}}, \bibinfo {author} {\bibfnamefont {D.}~\bibnamefont {Puntel}}, \bibinfo {author} {\bibfnamefont {F.}~\bibnamefont {Mazzola}}, \bibinfo {author} {\bibfnamefont {S.}~\bibnamefont {Fang}}, \bibinfo {author} {\bibfnamefont {C.}~\bibnamefont {Jozwiak}}, \bibinfo {author} {\bibfnamefont {A.}~\bibnamefont {Bostwick}}, \bibinfo {author} {\bibfnamefont {E.}~\bibnamefont {Rotenberg}}, \bibinfo {author} {\bibfnamefont {J.}~\bibnamefont {Fuji}}, \bibinfo {author} {\bibfnamefont {I.}~\bibnamefont {Vobornik}}, \bibinfo {author} {\bibfnamefont {J.-H.}\ \bibnamefont {Park}}, \bibinfo {author} {\bibfnamefont {J.~G.}\ \bibnamefont {Checkelsky}},
  \bibinfo {author} {\bibfnamefont {B.-J.}\ \bibnamefont {Yang}},\ and\ \bibinfo {author} {\bibfnamefont {R.}~\bibnamefont {Comin}},\ }\bibfield  {title} {\bibinfo {title} {Measurements of the quantum geometric tensor in solids},\ }\href {https://doi.org/10.1038/s41567-024-02678-8} {\bibfield  {journal} {\bibinfo  {journal} {Nat. Phys.}\ }\textbf {\bibinfo {volume} {21}},\ \bibinfo {pages} {110} (\bibinfo {year} {2025})}\BibitemShut {NoStop}%
\bibitem [{\citenamefont {Kim}\ \emph {et~al.}(2025)\citenamefont {Kim}, \citenamefont {Chung}, \citenamefont {Qian}, \citenamefont {Park}, \citenamefont {Jozwiak}, \citenamefont {Rotenberg}, \citenamefont {Bostwick}, \citenamefont {Kim},\ and\ \citenamefont {Yang}}]{Kim2025}%
  \BibitemOpen
  \bibfield  {author} {\bibinfo {author} {\bibfnamefont {S.}~\bibnamefont {Kim}}, \bibinfo {author} {\bibfnamefont {Y.}~\bibnamefont {Chung}}, \bibinfo {author} {\bibfnamefont {Y.}~\bibnamefont {Qian}}, \bibinfo {author} {\bibfnamefont {S.}~\bibnamefont {Park}}, \bibinfo {author} {\bibfnamefont {C.}~\bibnamefont {Jozwiak}}, \bibinfo {author} {\bibfnamefont {E.}~\bibnamefont {Rotenberg}}, \bibinfo {author} {\bibfnamefont {A.}~\bibnamefont {Bostwick}}, \bibinfo {author} {\bibfnamefont {K.~S.}\ \bibnamefont {Kim}},\ and\ \bibinfo {author} {\bibfnamefont {B.-J.}\ \bibnamefont {Yang}},\ }\bibfield  {title} {\bibinfo {title} {Direct measurement of the quantum metric tensor in solids},\ }\href {https://doi.org/10.1126/science.ado6049} {\bibfield  {journal} {\bibinfo  {journal} {Science}\ }\textbf {\bibinfo {volume} {388}},\ \bibinfo {pages} {1050} (\bibinfo {year} {2025})}\BibitemShut {NoStop}%
\bibitem [{\citenamefont {Shinada}\ and\ \citenamefont {Nagaosa}(2025)}]{Shinada2025}%
  \BibitemOpen
  \bibfield  {author} {\bibinfo {author} {\bibfnamefont {K.}~\bibnamefont {Shinada}}\ and\ \bibinfo {author} {\bibfnamefont {N.}~\bibnamefont {Nagaosa}},\ }\bibfield  {title} {\bibinfo {title} {{Quantum geometric bounds for observables: Linear responses, Drude weight, and orbital magnetization}},\ }\href {https://doi.org/10.1103/qxbl-qd4f} {\bibfield  {journal} {\bibinfo  {journal} {Phys. Rev. B}\ }\textbf {\bibinfo {volume} {112}},\ \bibinfo {pages} {155158} (\bibinfo {year} {2025})}\BibitemShut {NoStop}%
\bibitem [{\citenamefont {Jankowski}\ \emph {et~al.}(2025{\natexlab{d}})\citenamefont {Jankowski}, \citenamefont {Morris}, \citenamefont {Bouhon}, \citenamefont {\"Unal},\ and\ \citenamefont {Slager}}]{Jankowski2025PRBoptical}%
  \BibitemOpen
  \bibfield  {author} {\bibinfo {author} {\bibfnamefont {W.~J.}\ \bibnamefont {Jankowski}}, \bibinfo {author} {\bibfnamefont {A.~S.}\ \bibnamefont {Morris}}, \bibinfo {author} {\bibfnamefont {A.}~\bibnamefont {Bouhon}}, \bibinfo {author} {\bibfnamefont {F.~N.}\ \bibnamefont {\"Unal}},\ and\ \bibinfo {author} {\bibfnamefont {R.-J.}\ \bibnamefont {Slager}},\ }\bibfield  {title} {\bibinfo {title} {Optical manifestations and bounds of topological {E}uler class},\ }\href {https://doi.org/10.1103/PhysRevB.111.L081103} {\bibfield  {journal} {\bibinfo  {journal} {Phys. Rev. B}\ }\textbf {\bibinfo {volume} {111}},\ \bibinfo {pages} {L081103} (\bibinfo {year} {2025}{\natexlab{d}})}\BibitemShut {NoStop}%
\bibitem [{\citenamefont {Ji}\ \emph {et~al.}()\citenamefont {Ji}, \citenamefont {Palomino}, \citenamefont {Goldman}, \citenamefont {Ozawa}, \citenamefont {Riseborough}, \citenamefont {Wang},\ and\ \citenamefont {Mera}}]{Ji2025}%
  \BibitemOpen
  \bibfield  {author} {\bibinfo {author} {\bibfnamefont {G.}~\bibnamefont {Ji}}, \bibinfo {author} {\bibfnamefont {D.~E.}\ \bibnamefont {Palomino}}, \bibinfo {author} {\bibfnamefont {N.}~\bibnamefont {Goldman}}, \bibinfo {author} {\bibfnamefont {T.}~\bibnamefont {Ozawa}}, \bibinfo {author} {\bibfnamefont {P.}~\bibnamefont {Riseborough}}, \bibinfo {author} {\bibfnamefont {J.}~\bibnamefont {Wang}},\ and\ \bibinfo {author} {\bibfnamefont {B.}~\bibnamefont {Mera}},\ }\bibfield  {title} {\bibinfo {title} {Density matrix geometry and sum rules},\ }\Eprint {https://arxiv.org/abs/2507.14028} {arXiv:2507.14028} \BibitemShut {NoStop}%
\bibitem [{\citenamefont {Esin}\ \emph {et~al.}(2025)\citenamefont {Esin}, \citenamefont {Lantagne-Hurtubise}, \citenamefont {Nathan},\ and\ \citenamefont {Refael}}]{Esin2025}%
  \BibitemOpen
  \bibfield  {author} {\bibinfo {author} {\bibfnamefont {I.}~\bibnamefont {Esin}}, \bibinfo {author} {\bibfnamefont {E.}~\bibnamefont {Lantagne-Hurtubise}}, \bibinfo {author} {\bibfnamefont {F.}~\bibnamefont {Nathan}},\ and\ \bibinfo {author} {\bibfnamefont {G.}~\bibnamefont {Refael}},\ }\bibfield  {title} {\bibinfo {title} {Quantum geometry and bounds on dissipation in slowly driven quantum systems},\ }\href {https://doi.org/10.1103/PhysRevLett.134.146603} {\bibfield  {journal} {\bibinfo  {journal} {Phys. Rev. Lett.}\ }\textbf {\bibinfo {volume} {134}},\ \bibinfo {pages} {146603} (\bibinfo {year} {2025})}\BibitemShut {NoStop}%
\bibitem [{\citenamefont {Montag}\ and\ \citenamefont {Ozawa}(2026)}]{Montag2025}%
  \BibitemOpen
  \bibfield  {author} {\bibinfo {author} {\bibfnamefont {A.}~\bibnamefont {Montag}}\ and\ \bibinfo {author} {\bibfnamefont {T.}~\bibnamefont {Ozawa}},\ }\bibfield  {title} {\bibinfo {title} {{Quantum geometrical effects in non-Hermitian systems}},\ }\href {https://doi.org/10.1103/qb8s-9c6y} {\bibfield  {journal} {\bibinfo  {journal} {Phys. Rev. Res.}\ }\textbf {\bibinfo {volume} {8}},\ \bibinfo {pages} {013181} (\bibinfo {year} {2026})}\BibitemShut {NoStop}%
\bibitem [{\citenamefont {Singhal}\ \emph {et~al.}(2023)\citenamefont {Singhal}, \citenamefont {Martello}, \citenamefont {Agrawal}, \citenamefont {Ozawa}, \citenamefont {Price},\ and\ \citenamefont {Gadway}}]{Singhal:2023gl}%
  \BibitemOpen
  \bibfield  {author} {\bibinfo {author} {\bibfnamefont {Y.}~\bibnamefont {Singhal}}, \bibinfo {author} {\bibfnamefont {E.}~\bibnamefont {Martello}}, \bibinfo {author} {\bibfnamefont {S.}~\bibnamefont {Agrawal}}, \bibinfo {author} {\bibfnamefont {T.}~\bibnamefont {Ozawa}}, \bibinfo {author} {\bibfnamefont {H.}~\bibnamefont {Price}},\ and\ \bibinfo {author} {\bibfnamefont {B.}~\bibnamefont {Gadway}},\ }\bibfield  {title} {\bibinfo {title} {{Measuring the adiabatic non-Hermitian Berry phase in feedback-coupled oscillators}},\ }\href {https://doi.org/10.1103/PhysRevResearch.5.L032026} {\bibfield  {journal} {\bibinfo  {journal} {Phys. Rev. Res.}\ }\textbf {\bibinfo {volume} {5}},\ \bibinfo {pages} {L032026} (\bibinfo {year} {2023})}\BibitemShut {NoStop}%
\bibitem [{\citenamefont {Chen~Ye}\ \emph {et~al.}(2024)\citenamefont {Chen~Ye}, \citenamefont {Vleeshouwers}, \citenamefont {Heatley}, \citenamefont {Gritsev},\ and\ \citenamefont {Morais~Smith}}]{Ye2024}%
  \BibitemOpen
  \bibfield  {author} {\bibinfo {author} {\bibfnamefont {C.}~\bibnamefont {Chen~Ye}}, \bibinfo {author} {\bibfnamefont {W.~L.}\ \bibnamefont {Vleeshouwers}}, \bibinfo {author} {\bibfnamefont {S.}~\bibnamefont {Heatley}}, \bibinfo {author} {\bibfnamefont {V.}~\bibnamefont {Gritsev}},\ and\ \bibinfo {author} {\bibfnamefont {C.}~\bibnamefont {Morais~Smith}},\ }\bibfield  {title} {\bibinfo {title} {Quantum metric of non-{H}ermitian {S}u-{S}chrieffer-{H}eeger systems},\ }\href {https://doi.org/10.1103/PhysRevResearch.6.023202} {\bibfield  {journal} {\bibinfo  {journal} {Phys. Rev. Res.}\ }\textbf {\bibinfo {volume} {6}},\ \bibinfo {pages} {023202} (\bibinfo {year} {2024})}\BibitemShut {NoStop}%
\bibitem [{\citenamefont {Ozawa}\ and\ \citenamefont {Schomerus}(2025)}]{Ozawa:2025bv}%
  \BibitemOpen
  \bibfield  {author} {\bibinfo {author} {\bibfnamefont {T.}~\bibnamefont {Ozawa}}\ and\ \bibinfo {author} {\bibfnamefont {H.}~\bibnamefont {Schomerus}},\ }\bibfield  {title} {\bibinfo {title} {{Geometric contribution to adiabatic amplification in non-Hermitian systems}},\ }\href {https://doi.org/10.1103/PhysRevResearch.7.013173} {\bibfield  {journal} {\bibinfo  {journal} {Phys. Rev. Res.}\ }\textbf {\bibinfo {volume} {7}},\ \bibinfo {pages} {013173} (\bibinfo {year} {2025})}\BibitemShut {NoStop}%
\bibitem [{\citenamefont {Behrends}\ \emph {et~al.}()\citenamefont {Behrends}, \citenamefont {Ilan},\ and\ \citenamefont {Goldstein}}]{Behrends2025}%
  \BibitemOpen
  \bibfield  {author} {\bibinfo {author} {\bibfnamefont {J.}~\bibnamefont {Behrends}}, \bibinfo {author} {\bibfnamefont {R.}~\bibnamefont {Ilan}},\ and\ \bibinfo {author} {\bibfnamefont {M.}~\bibnamefont {Goldstein}},\ }\bibfield  {title} {\bibinfo {title} {Quantum geometry of non-{H}ermitian systems},\ }\Eprint {https://arxiv.org/abs/2503.13604} {arXiv:2503.13604} \BibitemShut {NoStop}%
\bibitem [{\citenamefont {Ashida}\ \emph {et~al.}(2020)\citenamefont {Ashida}, \citenamefont {Gong},\ and\ \citenamefont {Ueda}}]{Ashida:2020fo}%
  \BibitemOpen
  \bibfield  {author} {\bibinfo {author} {\bibfnamefont {Y.}~\bibnamefont {Ashida}}, \bibinfo {author} {\bibfnamefont {Z.}~\bibnamefont {Gong}},\ and\ \bibinfo {author} {\bibfnamefont {M.}~\bibnamefont {Ueda}},\ }\bibfield  {title} {\bibinfo {title} {{Non-Hermitian physics}},\ }\href {https://doi.org/10.1080/00018732.2021.1876991} {\bibfield  {journal} {\bibinfo  {journal} {Adv. Phys.}\ }\textbf {\bibinfo {volume} {69}},\ \bibinfo {pages} {249} (\bibinfo {year} {2020})}\BibitemShut {NoStop}%
\bibitem [{\citenamefont {Bergholtz}\ \emph {et~al.}(2021)\citenamefont {Bergholtz}, \citenamefont {Budich},\ and\ \citenamefont {Kunst}}]{Bergholtz:2021kc}%
  \BibitemOpen
  \bibfield  {author} {\bibinfo {author} {\bibfnamefont {E.~J.}\ \bibnamefont {Bergholtz}}, \bibinfo {author} {\bibfnamefont {J.~C.}\ \bibnamefont {Budich}},\ and\ \bibinfo {author} {\bibfnamefont {F.~K.}\ \bibnamefont {Kunst}},\ }\bibfield  {title} {\bibinfo {title} {{Exceptional topology of non-Hermitian systems}},\ }\href {https://doi.org/10.1103/RevModPhys.93.015005} {\bibfield  {journal} {\bibinfo  {journal} {Rev. Mod. Phys.}\ }\textbf {\bibinfo {volume} {93}},\ \bibinfo {pages} {015005} (\bibinfo {year} {2021})}\BibitemShut {NoStop}%
\bibitem [{\citenamefont {Herviou}\ \emph {et~al.}(2019)\citenamefont {Herviou}, \citenamefont {Bardarson},\ and\ \citenamefont {Regnault}}]{Herviou:2019ih}%
  \BibitemOpen
  \bibfield  {author} {\bibinfo {author} {\bibfnamefont {L.}~\bibnamefont {Herviou}}, \bibinfo {author} {\bibfnamefont {J.~H.}\ \bibnamefont {Bardarson}},\ and\ \bibinfo {author} {\bibfnamefont {N.}~\bibnamefont {Regnault}},\ }\bibfield  {title} {\bibinfo {title} {{Defining a bulk-edge correspondence for non-Hermitian Hamiltonians via singular-value decomposition}},\ }\href {https://doi.org/10.1103/PhysRevA.99.052118} {\bibfield  {journal} {\bibinfo  {journal} {Phys. Rev. A}\ }\textbf {\bibinfo {volume} {99}},\ \bibinfo {pages} {052118} (\bibinfo {year} {2019})}\BibitemShut {NoStop}%
\bibitem [{\citenamefont {Wanjura}\ \emph {et~al.}(2020)\citenamefont {Wanjura}, \citenamefont {Brunelli},\ and\ \citenamefont {Nunnenkamp}}]{Wanjura:2020jg}%
  \BibitemOpen
  \bibfield  {author} {\bibinfo {author} {\bibfnamefont {C.~C.}\ \bibnamefont {Wanjura}}, \bibinfo {author} {\bibfnamefont {M.}~\bibnamefont {Brunelli}},\ and\ \bibinfo {author} {\bibfnamefont {A.}~\bibnamefont {Nunnenkamp}},\ }\bibfield  {title} {\bibinfo {title} {{Topological framework for directional amplification in driven-dissipative cavity arrays}},\ }\href {https://doi.org/10.1038/s41467-020-16863-9} {\bibfield  {journal} {\bibinfo  {journal} {Nat. Comm.}\ }\textbf {\bibinfo {volume} {11}},\ \bibinfo {pages} {3149} (\bibinfo {year} {2020})}\BibitemShut {NoStop}%
\bibitem [{\citenamefont {Borgnia}\ \emph {et~al.}(2020)\citenamefont {Borgnia}, \citenamefont {Kruchkov},\ and\ \citenamefont {Slager}}]{Borgnia:2020hi}%
  \BibitemOpen
  \bibfield  {author} {\bibinfo {author} {\bibfnamefont {D.~S.}\ \bibnamefont {Borgnia}}, \bibinfo {author} {\bibfnamefont {A.~J.}\ \bibnamefont {Kruchkov}},\ and\ \bibinfo {author} {\bibfnamefont {R.-J.}\ \bibnamefont {Slager}},\ }\bibfield  {title} {\bibinfo {title} {{Non-Hermitian Boundary Modes and Topology}},\ }\href {https://doi.org/10.1103/PhysRevLett.124.056802} {\bibfield  {journal} {\bibinfo  {journal} {Phys. Rev. Lett.}\ }\textbf {\bibinfo {volume} {124}},\ \bibinfo {pages} {056802} (\bibinfo {year} {2020})}\BibitemShut {NoStop}%
\bibitem [{\citenamefont {Zirnstein}\ \emph {et~al.}(2021)\citenamefont {Zirnstein}, \citenamefont {Refael},\ and\ \citenamefont {Rosenow}}]{Zirnstein:2021cl}%
  \BibitemOpen
  \bibfield  {author} {\bibinfo {author} {\bibfnamefont {H.-G.}\ \bibnamefont {Zirnstein}}, \bibinfo {author} {\bibfnamefont {G.}~\bibnamefont {Refael}},\ and\ \bibinfo {author} {\bibfnamefont {B.}~\bibnamefont {Rosenow}},\ }\bibfield  {title} {\bibinfo {title} {{Bulk-Boundary Correspondence for Non-Hermitian Hamiltonians via Green Functions}},\ }\href {https://doi.org/10.1103/PhysRevLett.126.216407} {\bibfield  {journal} {\bibinfo  {journal} {Phys. Rev. Lett.}\ }\textbf {\bibinfo {volume} {126}},\ \bibinfo {pages} {216407} (\bibinfo {year} {2021})}\BibitemShut {NoStop}%
\bibitem [{\citenamefont {Shen}\ \emph {et~al.}(2018)\citenamefont {Shen}, \citenamefont {Zhen},\ and\ \citenamefont {Fu}}]{Shen2018}%
  \BibitemOpen
  \bibfield  {author} {\bibinfo {author} {\bibfnamefont {H.}~\bibnamefont {Shen}}, \bibinfo {author} {\bibfnamefont {B.}~\bibnamefont {Zhen}},\ and\ \bibinfo {author} {\bibfnamefont {L.}~\bibnamefont {Fu}},\ }\bibfield  {title} {\bibinfo {title} {{Topological Band Theory for Non-Hermitian Hamiltonians}},\ }\href {https://doi.org/10.1103/PhysRevLett.120.146402} {\bibfield  {journal} {\bibinfo  {journal} {Phys. Rev. Lett.}\ }\textbf {\bibinfo {volume} {120}},\ \bibinfo {pages} {146402} (\bibinfo {year} {2018})}\BibitemShut {NoStop}%
\bibitem [{\citenamefont {Gong}\ \emph {et~al.}(2018)\citenamefont {Gong}, \citenamefont {Ashida}, \citenamefont {Kawabata}, \citenamefont {Takasan}, \citenamefont {Higashikawa},\ and\ \citenamefont {Ueda}}]{Gong:2018ko}%
  \BibitemOpen
  \bibfield  {author} {\bibinfo {author} {\bibfnamefont {Z.}~\bibnamefont {Gong}}, \bibinfo {author} {\bibfnamefont {Y.}~\bibnamefont {Ashida}}, \bibinfo {author} {\bibfnamefont {K.}~\bibnamefont {Kawabata}}, \bibinfo {author} {\bibfnamefont {K.}~\bibnamefont {Takasan}}, \bibinfo {author} {\bibfnamefont {S.}~\bibnamefont {Higashikawa}},\ and\ \bibinfo {author} {\bibfnamefont {M.}~\bibnamefont {Ueda}},\ }\bibfield  {title} {\bibinfo {title} {{Topological Phases of Non-Hermitian Systems}},\ }\href {https://doi.org/10.1103/PhysRevX.8.031079} {\bibfield  {journal} {\bibinfo  {journal} {Phys. Rev. X}\ }\textbf {\bibinfo {volume} {8}},\ \bibinfo {pages} {031079} (\bibinfo {year} {2018})}\BibitemShut {NoStop}%
\bibitem [{\citenamefont {Prosen}(2008)}]{Prosen:2008dw}%
  \BibitemOpen
  \bibfield  {author} {\bibinfo {author} {\bibfnamefont {T.}~\bibnamefont {Prosen}},\ }\bibfield  {title} {\bibinfo {title} {{Third quantization: a general method to solve master equations for quadratic open Fermi systems}},\ }\href {https://doi.org/10.1088/1367-2630/10/4/043026} {\bibfield  {journal} {\bibinfo  {journal} {New J. Phys.}\ }\textbf {\bibinfo {volume} {10}},\ \bibinfo {pages} {043026} (\bibinfo {year} {2008})}\BibitemShut {NoStop}%
\bibitem [{\citenamefont {Prosen}(2010)}]{Prosen:2010hp}%
  \BibitemOpen
  \bibfield  {author} {\bibinfo {author} {\bibfnamefont {T.}~\bibnamefont {Prosen}},\ }\bibfield  {title} {\bibinfo {title} {{Spectral theorem for the Lindblad equation for quadratic open fermionic systems}},\ }\href {https://doi.org/10.1088/1742-5468/2010/07/P07020} {\bibfield  {journal} {\bibinfo  {journal} {J. Stat. Mech.: Theory Exp.}\ }\textbf {\bibinfo {volume} {2010}}\bibinfo  {number} { (07)},\ \bibinfo {pages} {P07020}}\BibitemShut {NoStop}%
\bibitem [{\citenamefont {Kos}\ and\ \citenamefont {Prosen}(2017)}]{Kos:2017ew}%
  \BibitemOpen
\bibfield  {number} {  }\bibfield  {author} {\bibinfo {author} {\bibfnamefont {P.}~\bibnamefont {Kos}}\ and\ \bibinfo {author} {\bibfnamefont {T.}~\bibnamefont {Prosen}},\ }\bibfield  {title} {\bibinfo {title} {{Time-dependent correlation functions in open quadratic fermionic systems}},\ }\href {https://doi.org/10.1088/1742-5468/aa9681} {\bibfield  {journal} {\bibinfo  {journal} {J. Stat. Mech.: Theory Exp.}\ }\textbf {\bibinfo {volume} {2017}}\bibinfo  {number} { (12)},\ \bibinfo {pages} {123103}}\BibitemShut {NoStop}%
\bibitem [{\citenamefont {Lieu}\ \emph {et~al.}(2020)\citenamefont {Lieu}, \citenamefont {McGinley},\ and\ \citenamefont {Cooper}}]{Lieu:2020bc}%
  \BibitemOpen
\bibfield  {number} {  }\bibfield  {author} {\bibinfo {author} {\bibfnamefont {S.}~\bibnamefont {Lieu}}, \bibinfo {author} {\bibfnamefont {M.}~\bibnamefont {McGinley}},\ and\ \bibinfo {author} {\bibfnamefont {N.~R.}\ \bibnamefont {Cooper}},\ }\bibfield  {title} {\bibinfo {title} {{Tenfold Way for Quadratic Lindbladians}},\ }\href {https://doi.org/10.1103/PhysRevLett.124.040401} {\bibfield  {journal} {\bibinfo  {journal} {Phys. Rev. Lett.}\ }\textbf {\bibinfo {volume} {124}},\ \bibinfo {pages} {040401} (\bibinfo {year} {2020})}\BibitemShut {NoStop}%
\bibitem [{\citenamefont {S{\'{a}}}\ \emph {et~al.}(2023)\citenamefont {S{\'{a}}}, \citenamefont {Ribeiro},\ and\ \citenamefont {Prosen}}]{Sa:2023go}%
  \BibitemOpen
  \bibfield  {author} {\bibinfo {author} {\bibfnamefont {L.}~\bibnamefont {S{\'{a}}}}, \bibinfo {author} {\bibfnamefont {P.}~\bibnamefont {Ribeiro}},\ and\ \bibinfo {author} {\bibfnamefont {T.}~\bibnamefont {Prosen}},\ }\bibfield  {title} {\bibinfo {title} {{Symmetry Classification of Many-Body Lindbladians: Tenfold Way and Beyond}},\ }\href {https://doi.org/10.1103/PhysRevX.13.031019} {\bibfield  {journal} {\bibinfo  {journal} {Phys. Rev. X}\ }\textbf {\bibinfo {volume} {13}},\ \bibinfo {pages} {031019} (\bibinfo {year} {2023})}\BibitemShut {NoStop}%
\bibitem [{\citenamefont {Kawabata}\ \emph {et~al.}(2023)\citenamefont {Kawabata}, \citenamefont {Kulkarni}, \citenamefont {Li}, \citenamefont {Numasawa},\ and\ \citenamefont {Ryu}}]{Kawabata:2023hk}%
  \BibitemOpen
  \bibfield  {author} {\bibinfo {author} {\bibfnamefont {K.}~\bibnamefont {Kawabata}}, \bibinfo {author} {\bibfnamefont {A.}~\bibnamefont {Kulkarni}}, \bibinfo {author} {\bibfnamefont {J.}~\bibnamefont {Li}}, \bibinfo {author} {\bibfnamefont {T.}~\bibnamefont {Numasawa}},\ and\ \bibinfo {author} {\bibfnamefont {S.}~\bibnamefont {Ryu}},\ }\bibfield  {title} {\bibinfo {title} {{Symmetry of Open Quantum Systems: Classification of Dissipative Quantum Chaos}},\ }\href {https://doi.org/10.1103/PRXQuantum.4.030328} {\bibfield  {journal} {\bibinfo  {journal} {PRX Quantum}\ }\textbf {\bibinfo {volume} {4}},\ \bibinfo {pages} {030328} (\bibinfo {year} {2023})}\BibitemShut {NoStop}%
\bibitem [{\citenamefont {Xu}\ \emph {et~al.}(2017)\citenamefont {Xu}, \citenamefont {Wang},\ and\ \citenamefont {Duan}}]{Xu:2017bl}%
  \BibitemOpen
  \bibfield  {author} {\bibinfo {author} {\bibfnamefont {Y.}~\bibnamefont {Xu}}, \bibinfo {author} {\bibfnamefont {S.-T.}\ \bibnamefont {Wang}},\ and\ \bibinfo {author} {\bibfnamefont {L.-M.}\ \bibnamefont {Duan}},\ }\bibfield  {title} {\bibinfo {title} {{Weyl Exceptional Rings in a Three-Dimensional Dissipative Cold Atomic Gas}},\ }\href {https://doi.org/10.1103/PhysRevLett.118.045701} {\bibfield  {journal} {\bibinfo  {journal} {Phys. Rev. Lett.}\ }\textbf {\bibinfo {volume} {118}},\ \bibinfo {pages} {045701} (\bibinfo {year} {2017})}\BibitemShut {NoStop}%
\bibitem [{\citenamefont {Silberstein}\ \emph {et~al.}(2020)\citenamefont {Silberstein}, \citenamefont {Behrends}, \citenamefont {Goldstein},\ and\ \citenamefont {Ilan}}]{Silberstein:2020hi}%
  \BibitemOpen
  \bibfield  {author} {\bibinfo {author} {\bibfnamefont {N.}~\bibnamefont {Silberstein}}, \bibinfo {author} {\bibfnamefont {J.}~\bibnamefont {Behrends}}, \bibinfo {author} {\bibfnamefont {M.}~\bibnamefont {Goldstein}},\ and\ \bibinfo {author} {\bibfnamefont {R.}~\bibnamefont {Ilan}},\ }\bibfield  {title} {\bibinfo {title} {{Berry connection induced anomalous wave-packet dynamics in non-Hermitian systems}},\ }\href {https://doi.org/10.1103/PhysRevB.102.245147} {\bibfield  {journal} {\bibinfo  {journal} {Phys. Rev. B}\ }\textbf {\bibinfo {volume} {102}},\ \bibinfo {pages} {245147} (\bibinfo {year} {2020})}\BibitemShut {NoStop}%
\bibitem [{\citenamefont {Hu}\ \emph {et~al.}(2025)\citenamefont {Hu}, \citenamefont {Ostrovskaya},\ and\ \citenamefont {Estrecho}}]{Hu:2025ct}%
  \BibitemOpen
  \bibfield  {author} {\bibinfo {author} {\bibfnamefont {Y.-M.~R.}\ \bibnamefont {Hu}}, \bibinfo {author} {\bibfnamefont {E.~A.}\ \bibnamefont {Ostrovskaya}},\ and\ \bibinfo {author} {\bibfnamefont {E.}~\bibnamefont {Estrecho}},\ }\bibfield  {title} {\bibinfo {title} {{Quantum geometric tensor and wavepacket dynamics in two-dimensional non-Hermitian systems}},\ }\href {https://doi.org/10.1103/PhysRevResearch.7.L012067} {\bibfield  {journal} {\bibinfo  {journal} {Phys. Rev. Res.}\ }\textbf {\bibinfo {volume} {7}},\ \bibinfo {pages} {L012067} (\bibinfo {year} {2025})}\BibitemShut {NoStop}%
\bibitem [{\citenamefont {Rice}\ and\ \citenamefont {Mele}(1982)}]{Rice:1982cr}%
  \BibitemOpen
  \bibfield  {author} {\bibinfo {author} {\bibfnamefont {M.~J.}\ \bibnamefont {Rice}}\ and\ \bibinfo {author} {\bibfnamefont {E.~J.}\ \bibnamefont {Mele}},\ }\bibfield  {title} {\bibinfo {title} {{Elementary Excitations of a Linearly Conjugated Diatomic Polymer}},\ }\href {https://doi.org/10.1103/PhysRevLett.49.1455} {\bibfield  {journal} {\bibinfo  {journal} {Phys. Rev. Lett.}\ }\textbf {\bibinfo {volume} {49}},\ \bibinfo {pages} {1455} (\bibinfo {year} {1982})}\BibitemShut {NoStop}%
\bibitem [{\citenamefont {Lin}\ \emph {et~al.}(2015)\citenamefont {Lin}, \citenamefont {Zhang},\ and\ \citenamefont {Song}}]{Lin:2015jk}%
  \BibitemOpen
  \bibfield  {author} {\bibinfo {author} {\bibfnamefont {S.}~\bibnamefont {Lin}}, \bibinfo {author} {\bibfnamefont {X.~Z.}\ \bibnamefont {Zhang}},\ and\ \bibinfo {author} {\bibfnamefont {Z.}~\bibnamefont {Song}},\ }\bibfield  {title} {\bibinfo {title} {{Amplitude control of a quantum state in a non-Hermitian Rice-Mele model driven by an external field}},\ }\href {https://doi.org/10.1103/PhysRevA.92.012117} {\bibfield  {journal} {\bibinfo  {journal} {Phys. Rev. A}\ }\textbf {\bibinfo {volume} {92}},\ \bibinfo {pages} {012117} (\bibinfo {year} {2015})}\BibitemShut {NoStop}%
\bibitem [{\citenamefont {Kunst}\ \emph {et~al.}(2018)\citenamefont {Kunst}, \citenamefont {Edvardsson}, \citenamefont {Budich},\ and\ \citenamefont {Bergholtz}}]{Kunst2018}%
  \BibitemOpen
  \bibfield  {author} {\bibinfo {author} {\bibfnamefont {F.~K.}\ \bibnamefont {Kunst}}, \bibinfo {author} {\bibfnamefont {E.}~\bibnamefont {Edvardsson}}, \bibinfo {author} {\bibfnamefont {J.~C.}\ \bibnamefont {Budich}},\ and\ \bibinfo {author} {\bibfnamefont {E.~J.}\ \bibnamefont {Bergholtz}},\ }\bibfield  {title} {\bibinfo {title} {{Biorthogonal Bulk-Boundary Correspondence in Non-Hermitian Systems}},\ }\href {https://doi.org/10.1103/PhysRevLett.121.026808} {\bibfield  {journal} {\bibinfo  {journal} {Phys. Rev. Lett.}\ }\textbf {\bibinfo {volume} {121}},\ \bibinfo {pages} {026808} (\bibinfo {year} {2018})}\BibitemShut {NoStop}%
\bibitem [{\citenamefont {Kamenev}(2011)}]{Kamenev_2011}%
  \BibitemOpen
  \bibfield  {author} {\bibinfo {author} {\bibfnamefont {A.}~\bibnamefont {Kamenev}},\ }\href {https://doi.org/10.1017/CBO9781139003667} {\emph {\bibinfo {title} {{Field Theory of Non-Equilibrium Systems}}}}\ (\bibinfo  {publisher} {Cambridge University Press},\ \bibinfo {address} {Cambridge, U.K.},\ \bibinfo {year} {2011})\BibitemShut {NoStop}%
\bibitem [{\citenamefont {Sieberer}\ \emph {et~al.}(2016)\citenamefont {Sieberer}, \citenamefont {Buchhold},\ and\ \citenamefont {Diehl}}]{Sieberer:2016ej}%
  \BibitemOpen
  \bibfield  {author} {\bibinfo {author} {\bibfnamefont {L.~M.}\ \bibnamefont {Sieberer}}, \bibinfo {author} {\bibfnamefont {M.}~\bibnamefont {Buchhold}},\ and\ \bibinfo {author} {\bibfnamefont {S.}~\bibnamefont {Diehl}},\ }\bibfield  {title} {\bibinfo {title} {{Keldysh field theory for driven open quantum systems}},\ }\href {https://doi.org/10.1088/0034-4885/79/9/096001} {\bibfield  {journal} {\bibinfo  {journal} {Rep. Prog. Phys.}\ }\textbf {\bibinfo {volume} {79}},\ \bibinfo {pages} {096001} (\bibinfo {year} {2016})}\BibitemShut {NoStop}%
\bibitem [{\citenamefont {Talkington}\ and\ \citenamefont {Claassen}(2024)}]{Talkington:2024gf}%
  \BibitemOpen
  \bibfield  {author} {\bibinfo {author} {\bibfnamefont {S.}~\bibnamefont {Talkington}}\ and\ \bibinfo {author} {\bibfnamefont {M.}~\bibnamefont {Claassen}},\ }\bibfield  {title} {\bibinfo {title} {{Linear and non-linear response of quadratic Lindbladians}},\ }\href {https://doi.org/10.1038/s41535-024-00709-4} {\bibfield  {journal} {\bibinfo  {journal} {npj Quantum Mater.}\ }\textbf {\bibinfo {volume} {9}},\ \bibinfo {pages} {104} (\bibinfo {year} {2024})}\BibitemShut {NoStop}%
\bibitem [{\citenamefont {Sieberer}\ \emph {et~al.}(2025)\citenamefont {Sieberer}, \citenamefont {Buchhold}, \citenamefont {Marino},\ and\ \citenamefont {Diehl}}]{Sieberer:2025eo}%
  \BibitemOpen
  \bibfield  {author} {\bibinfo {author} {\bibfnamefont {L.~M.}\ \bibnamefont {Sieberer}}, \bibinfo {author} {\bibfnamefont {M.}~\bibnamefont {Buchhold}}, \bibinfo {author} {\bibfnamefont {J.}~\bibnamefont {Marino}},\ and\ \bibinfo {author} {\bibfnamefont {S.}~\bibnamefont {Diehl}},\ }\bibfield  {title} {\bibinfo {title} {{Universality in driven open quantum matter}},\ }\href {https://doi.org/10.1103/RevModPhys.97.025004} {\bibfield  {journal} {\bibinfo  {journal} {Rev. Mod. Phys.}\ }\textbf {\bibinfo {volume} {97}},\ \bibinfo {pages} {025004} (\bibinfo {year} {2025})}\BibitemShut {NoStop}%
\bibitem [{\citenamefont {Weimer}\ \emph {et~al.}(2021)\citenamefont {Weimer}, \citenamefont {Kshetrimayum},\ and\ \citenamefont {Or{\'{u}}s}}]{Weimer:2021kp}%
  \BibitemOpen
  \bibfield  {author} {\bibinfo {author} {\bibfnamefont {H.}~\bibnamefont {Weimer}}, \bibinfo {author} {\bibfnamefont {A.}~\bibnamefont {Kshetrimayum}},\ and\ \bibinfo {author} {\bibfnamefont {R.}~\bibnamefont {Or{\'{u}}s}},\ }\bibfield  {title} {\bibinfo {title} {{Simulation methods for open quantum many-body systems}},\ }\href {https://doi.org/10.1103/RevModPhys.93.015008} {\bibfield  {journal} {\bibinfo  {journal} {Reviews of Modern Physics}\ }\textbf {\bibinfo {volume} {93}},\ \bibinfo {pages} {015008} (\bibinfo {year} {2021})}\BibitemShut {NoStop}%
\bibitem [{\citenamefont {Altland}\ and\ \citenamefont {Simons}(2010)}]{AltlandSimons2010}%
  \BibitemOpen
  \bibfield  {author} {\bibinfo {author} {\bibfnamefont {A.}~\bibnamefont {Altland}}\ and\ \bibinfo {author} {\bibfnamefont {B.}~\bibnamefont {Simons}},\ }\href {https://doi.org/10.1017/CBO9780511789984} {\emph {\bibinfo {title} {Condensed Matter Field Theory}}},\ \bibinfo {edition} {2nd}\ ed.\ (\bibinfo  {publisher} {Cambridge University Press},\ \bibinfo {address} {Cambridge},\ \bibinfo {year} {2010})\BibitemShut {NoStop}%
\bibitem [{\citenamefont {Brody}(2014)}]{Brody:2014jv}%
  \BibitemOpen
  \bibfield  {author} {\bibinfo {author} {\bibfnamefont {D.~C.}\ \bibnamefont {Brody}},\ }\bibfield  {title} {\bibinfo {title} {{Biorthogonal quantum mechanics}},\ }\href {https://doi.org/10.1088/1751-8113/47/3/035305} {\bibfield  {journal} {\bibinfo  {journal} {J. Phys. A: Math. Theor.}\ }\textbf {\bibinfo {volume} {47}},\ \bibinfo {pages} {035305} (\bibinfo {year} {2014})}\BibitemShut {NoStop}%
\bibitem [{Sup()}]{Supplemental}%
  \BibitemOpen
  \href@noop {} {}\bibinfo {note} {{See the Supplemental Material (SM) for further details and derivations of non-Hermitian quantum geometry and bounds (Sec.~{I}), and non-Hermitian geometric bounds in responses of open quantum systems (Sec.~{II}). The Supplemental Material also contains Refs. \cite{Kudinov1991, Daley:2014hc, Monkman2025}.}}\BibitemShut {Stop}%
\bibitem [{\citenamefont {Souza}\ \emph {et~al.}(2000)\citenamefont {Souza}, \citenamefont {Wilkens},\ and\ \citenamefont {Martin}}]{Souza:2000cj}%
  \BibitemOpen
  \bibfield  {author} {\bibinfo {author} {\bibfnamefont {I.}~\bibnamefont {Souza}}, \bibinfo {author} {\bibfnamefont {T.}~\bibnamefont {Wilkens}},\ and\ \bibinfo {author} {\bibfnamefont {R.~M.}\ \bibnamefont {Martin}},\ }\bibfield  {title} {\bibinfo {title} {{Polarization and localization in insulators: Generating function approach}},\ }\href {https://doi.org/10.1103/PhysRevB.62.1666} {\bibfield  {journal} {\bibinfo  {journal} {Phys. Rev. B}\ }\textbf {\bibinfo {volume} {62}},\ \bibinfo {pages} {1666} (\bibinfo {year} {2000})}\BibitemShut {NoStop}%
\bibitem [{\citenamefont {Manzano}(2020)}]{Manzano:2020eb}%
  \BibitemOpen
  \bibfield  {author} {\bibinfo {author} {\bibfnamefont {D.}~\bibnamefont {Manzano}},\ }\bibfield  {title} {\bibinfo {title} {{A short introduction to the Lindblad master equation}},\ }\href {https://doi.org/10.1063/1.5115323} {\bibfield  {journal} {\bibinfo  {journal} {AIP Adv.}\ }\textbf {\bibinfo {volume} {10}},\ \bibinfo {pages} {025106} (\bibinfo {year} {2020})}\BibitemShut {NoStop}%
\bibitem [{\citenamefont {Zhu}\ \emph {et~al.}(2021)\citenamefont {Zhu}, \citenamefont {Zheng}, \citenamefont {Zhu},\ and\ \citenamefont {Palumbo}}]{Zhu2021}%
  \BibitemOpen
  \bibfield  {author} {\bibinfo {author} {\bibfnamefont {Y.-Q.}\ \bibnamefont {Zhu}}, \bibinfo {author} {\bibfnamefont {W.}~\bibnamefont {Zheng}}, \bibinfo {author} {\bibfnamefont {S.-L.}\ \bibnamefont {Zhu}},\ and\ \bibinfo {author} {\bibfnamefont {G.}~\bibnamefont {Palumbo}},\ }\bibfield  {title} {\bibinfo {title} {{Band topology of pseudo-Hermitian phases through tensor Berry connections and quantum metric}},\ }\href {https://doi.org/10.1103/PhysRevB.104.205103} {\bibfield  {journal} {\bibinfo  {journal} {Phys. Rev. B}\ }\textbf {\bibinfo {volume} {104}},\ \bibinfo {pages} {205103} (\bibinfo {year} {2021})}\BibitemShut {NoStop}%
\bibitem [{\citenamefont {del Pino}\ \emph {et~al.}(2022)\citenamefont {del Pino}, \citenamefont {Slim},\ and\ \citenamefont {Verhagen}}]{delPino2022}%
  \BibitemOpen
  \bibfield  {author} {\bibinfo {author} {\bibfnamefont {J.}~\bibnamefont {del Pino}}, \bibinfo {author} {\bibfnamefont {J.~J.}\ \bibnamefont {Slim}},\ and\ \bibinfo {author} {\bibfnamefont {E.}~\bibnamefont {Verhagen}},\ }\bibfield  {title} {\bibinfo {title} {Non-{H}ermitian chiral phononics through optomechanically induced squeezing},\ }\href {https://doi.org/10.1038/s41586-022-04609-0} {\bibfield  {journal} {\bibinfo  {journal} {Nature}\ }\textbf {\bibinfo {volume} {606}},\ \bibinfo {pages} {82} (\bibinfo {year} {2022})}\BibitemShut {NoStop}%
\bibitem [{\citenamefont {Kudinov}(1991)}]{Kudinov1991}%
  \BibitemOpen
  \bibfield  {author} {\bibinfo {author} {\bibfnamefont {E.~K.}\ \bibnamefont {Kudinov}},\ }\bibfield  {title} {\bibinfo {title} {{Difference between insulating and conducting states}},\ }\href@noop {} {\bibfield  {journal} {\bibinfo  {journal} {Fiz. Tverd. Tela}\ }\textbf {\bibinfo {volume} {33}},\ \bibinfo {pages} {2306} (\bibinfo {year} {1991})},\ \bibinfo {note} {[Sov.\ Phys.\ Solid State 33, 1299 (1991)]}\BibitemShut {NoStop}%
\bibitem [{\citenamefont {Daley}(2014)}]{Daley:2014hc}%
  \BibitemOpen
  \bibfield  {author} {\bibinfo {author} {\bibfnamefont {A.~J.}\ \bibnamefont {Daley}},\ }\bibfield  {title} {\bibinfo {title} {{Quantum trajectories and open many-body quantum systems}},\ }\href {https://doi.org/10.1080/00018732.2014.933502} {\bibfield  {journal} {\bibinfo  {journal} {Advances in Physics}\ }\textbf {\bibinfo {volume} {63}},\ \bibinfo {pages} {77} (\bibinfo {year} {2014})}\BibitemShut {NoStop}%
\bibitem [{\citenamefont {Monkman}\ and\ \citenamefont {Berciu}(2026)}]{Monkman2025}%
  \BibitemOpen
  \bibfield  {author} {\bibinfo {author} {\bibfnamefont {K.}~\bibnamefont {Monkman}}\ and\ \bibinfo {author} {\bibfnamefont {M.}~\bibnamefont {Berciu}},\ }\bibfield  {title} {\bibinfo {title} {{Limits of the non-Hermitian description of decay models}},\ }\href {https://doi.org/10.1103/dyjw-1cbl} {\bibfield  {journal} {\bibinfo  {journal} {Phys. Rev. A}\ }\textbf {\bibinfo {volume} {113}},\ \bibinfo {pages} {032213} (\bibinfo {year} {2026})}\BibitemShut {NoStop}%
\end{thebibliography}%

\end{document}


\title{SUPPLEMENTAL MATERIAL \\ Quantum geometric bounds in non-Hermitian systems}

\newcommand{\TCM}{{TCM Group, Cavendish Laboratory, University of Cambridge, J.\,J.\,Thomson Avenue, Cambridge CB3 0HE, UK}}
\newcommand{\MPIPKS}{{Max-Planck-Institut f\"ur Physik komplexer Systeme, N\"othnitzer Stra\ss e 38, 01187 Dresden, Germany}}
\newcommand{\quantinuum}{Quantinuum, Terrington House, 13-15 Hills Rd, Cambridge, CB2 1NL, UK}

\author{Milosz Matraszek}
\email{mpm70@cam.ac.uk}
\affiliation{\TCM}
\affiliation{\MPIPKS}

\author{Wojciech J. Jankowski}
\email{wjj25@cam.ac.uk}
\affiliation{\TCM}
\affiliation{\MPIPKS}

\author{Jan Behrends}
\email{jan.behrends@yahoo.de}
\altaffiliation[Present address: ]{\quantinuum}
\affiliation{\TCM}

\date{\today}

\maketitle

\tableofcontents

\newpage

\section{Non-Hermitian quantum geometry and bounds}\label{app::A}

In the following, we provide technical details on the non-Hermitian quantum geometry and derivations of non-Hermitian quantum bounds introduced in this work.

\subsection{Positive semidefiniteness of non-Hermitian quantum geometric tensors}\label{QGT-symmetric-positivesemidefinitive}

We first prove that the symmetric non-Hermitian quantum geometric tensor (QGT) is a positive semidefinite matrix. Correspondingly, we use a similar argument as for the Hermitian case~\cite{Torma2023}. We consider a matrix of a quadratic form, applied to an arbitrary complex vector: 
\begin{equation}
    v_\mu^*Q^{\alpha\alpha}_{\mu\nu}v_\nu = v_\mu^*\frac{\bra{\partial_\mu\psi^\alpha}(1-P^{\alpha\alpha})\ket{\partial_{\nu_{}}\psi^\alpha}}{\braket{\psi^\alpha}{\psi^\alpha}}v_\nu = v_\mu^*\frac{\bra{\partial_\mu\psi^\alpha}(1-P^{\alpha\alpha})^2\ket{\partial_{\nu_{}}\psi^\alpha}}{\braket{\psi^\alpha}{\psi^\alpha}}v_\nu,
\end{equation}
where $\alpha\in\{ \text{R},\text{L}\}$ is the basis index and $P^{\alpha\alpha} = \frac{\ket{\psi^\alpha}\bra{\psi^\alpha}}{\braket{\psi^\alpha}{\psi^\alpha}}$ is a projector operator. In the second equality, we have used $\left(P^{\alpha\alpha}\right)^2=P^{\alpha\alpha}$. Since $P^{\alpha\alpha}$ is Hermitian,
\begin{align*}
     v_\mu^*Q^{\alpha\alpha}_{\mu\nu}v_\nu &= \frac{1}{\braket{\psi^\alpha}{\psi^\alpha}}\left[(1-P^{\alpha\alpha})v_\mu\partial_\mu\ket{\psi^\alpha}\right]^\dagger\left[(1-P^{\alpha\alpha})v_\nu\partial_\nu\ket{\psi^\alpha}\right] \\
     &= \frac{1}{\braket{\psi^\alpha}{\psi^\alpha}}\left[(1-P^{\alpha\alpha})\mathbf{v}\cdot\nabla\ket{\psi^\alpha}\right]^\dagger\left[(1-P^{\alpha\alpha})\mathbf{v}\cdot\nabla\ket{\psi^\alpha}\right] \\
     &= \frac{||(1-P^{\alpha\alpha})\mathbf{v}\cdot\nabla\ket{\psi^\alpha}||^2}{\braket{\psi^\alpha}{\psi^\alpha}} \geqslant 0,
\end{align*}
where the last inequality follows from the positivity of the norm.

\subsection{Non-Hermitian quantum geometric tensor bounds}

Below, we derive quantum geometric bounds satisfied with non-Hermitian quantum QGT. We consider the non-Hermitian QGT elements:
\beq{}
   Q^{\text{LR}}_{ij} = \frac{\bra{\partial_{k_i} \psi^{\text{L}}_n} \Big(1 - \ket{\psi^{\text{R}}_n} \bra{\psi^{\text{L}}_n}\Big) \ket{\partial_{k_j} \psi^{\text{R}}_n}}{\bra{\psi^{\text{L}}_n} \ket{ \psi^{\text{L}}_n}\bra{\psi^{\text{R}}_n } \ket{\psi^{\text{R}}_n}}.
\eeq
Notably, the mixed (left-right) LR/RL QGT is a non-Hermitian matrix and fails to be positive semidefinite, unlike the RR and LL QGTs. As such, we show that the following inequality holds:
\begin{equation}
    |Q^{RL}_{\mu\nu}|^2 \leqslant ||\psi^{\text{R}}_n||^2||\psi^{\text{L}}_n||^2\left(Q^{\text{RR}}_{\mu\mu} + |Q^{\text{R}}_\mu|^2 \right)\left( Q^{\text{LL}}_{\nu\nu} + |Q^{\text{L}}_\nu|^2 \right).
    \label{QGT-inequality}
\end{equation}
To prove this geometric bound, we define auxiliary states $|a\rangle$, $|b\rangle$:
\begin{align}
    |a\rangle &= \big(1-|\psi_n^{\text{L}}\rangle\langle \psi_n^{\text{R}}|\big)|\partial_\nu \psi_n^{\text{L}}\rangle, \\
    |b\rangle &= \big(1-|\psi_n^{\text{R}}\rangle\langle \psi_n^{\text{L}}|\big)|\partial_\mu \psi_n^{\text{R}}\rangle.
\end{align}
Correspondingly, we compute the inner products,
\begin{align*}
    \braket{a}{a} =& \bra{\partial_\nu\psi^{\text{L}}_n}(1-\ket{\psi^{\text{R}}_n}\bra{\psi^{\text{L}}_n})(1-\ket{\psi^{\text{L}}_n}\bra{\psi^{\text{R}}_n})\ket{\partial_\nu\psi^{\text{L}}_n} \\
    =& \bra{\partial_\nu\psi^{\text{L}}_n}\left[1+||\psi^{\text{L}}_n||^2\ket{\psi^{\text{R}}_n}\bra{\psi^{\text{R}}_n}-\ket{\psi^{\text{L}}_n}\bra{\psi^{\text{R}}_n}-\ket{\psi^{\text{R}}_n}\bra{\psi^{\text{L}}_n}\right]\ket{\partial_\nu\psi^{\text{L}}_n} \\
    =& \bra{\partial_\nu\psi^{\text{L}}_n}\left[1-\frac{\ket{\psi^{\text{L}}_n}\bra{\psi^{\text{L}}_n}}{\braket{\psi^{\text{L}}_n}{\psi^{\text{L}}_n}}+\frac{\ket{\psi^{\text{L}}_n}\bra{\psi^{\text{L}}_n}}{\braket{\psi^{\text{L}}_n}{\psi^{\text{L}}_n}}+||\psi^{\text{L}}_n||^2\ket{\psi^{\text{R}}_n}\bra{\psi^{\text{R}}_n}-\ket{\psi^{\text{L}}_n}\bra{\psi^{\text{R}}_n}-\ket{\psi^{\text{R}}_n}\bra{\psi^{\text{L}}_n}\right]\ket{\partial_\nu\psi^{\text{L}}_n} \\
    =& ||\psi^{\text{L}}_n||^2 \left(\frac{\bra{\partial_\nu\psi^{\text{L}}_n}(1-P^{\text{LL}}_n)\ket{\partial_\nu\psi^{\text{L}}_n}}{\braket{\psi^{\text{L}}_n}{\psi^{\text{L}}_n}} + \frac{\braket{\partial_\nu\psi^{\text{L}}_n}{\psi^{\text{L}}_n}\braket{\psi^{\text{L}}_n}{\partial_\nu\psi^{\text{L}}_n}}{\braket{\psi^{\text{L}}_n|\psi^{\text{L}}_n}^2} - \frac{\braket{\partial_\nu\psi^{\text{L}}_n}{\psi^{\text{L}}_n}\braket{\psi^{\text{R}}_n}{\partial_\nu\psi^{\text{L}}_n}}{\braket{\psi^{\text{L}}_n}{\psi^{\text{L}}_n}}\right. \\
    &\left.-\left[\frac{\braket{\partial_\nu\psi^{\text{L}}_n}{\psi^{\text{L}}_n}\braket{\psi^{\text{R}}_n}{\partial_\nu\psi^{\text{L}}_n}}{\braket{\psi^{\text{L}}_n}{\psi^{\text{L}}_n}}\right]^* +\braket{\partial_\nu\psi^{\text{L}}_n}{\psi^{\text{R}}_n}\braket{\psi^{\text{R}}_n}{\partial_\nu\psi^{\text{L}}_n} \right) \\
    =& ||\psi^{\text{L}}_n||^2 \left(Q^{\text{LL}}_{\nu\nu} + \left|A^{\text{LL}}_{\nu}\right|^2 + \left|A^{\text{RL}}_{\nu}\right|^2 -2\Re\left(A^{\text{LL}}_{\nu}A^{\text{RL}}_{\nu}\right) \right) \\
    =& ||\psi^{\text{L}}_n||^2 \left(Q^{\text{LL}}_{\nu\nu} +  \left|Q^{\text{L}}_{\nu}\right|^2\right).
\end{align*} 
The formula for $\braket{b}{b}$ follows by mapping: $i\mapsto j,\,R\leftrightarrow L$. Similarly:
\begin{equation}
    \braket{b}{a} =\bra{\partial_\mu\psi^{\text{R}}_n}(1-\ket{\psi^{\text{L}}_n}\bra{\psi^{\text{R}}_n})(1-\ket{\psi^{\text{L}}_n}\bra{\psi^{\text{R}}_n})\ket{\partial_\nu\psi^{\text{L}}_n} = \bra{\partial_\mu\psi^{\text{R}}_n}(1-\ket{\psi^{\text{L}}_n}\bra{\psi^{\text{R}}_n})\ket{\partial_\nu\psi^{\text{L}}_n} = Q^{\text{RL}}_{\mu\nu},
\end{equation}
{Thus the inequality follows from the Cauchy--Schwarz inequality applied to the states $\ket{a}$ and $\ket{b}$, see Eq.~\eqref{QGT-inequality}.} We note that $||\psi^{\text{L}}_n||^2 ||\psi^{\text{R}}_n||^2$ present in the bound is a gauge-invariant quantity.

\subsection{Non-Hermitian Chern number bounds}\label{NHChern}

We further provide derivations of geometric bounds due to non-Hermitian Chern numbers ($C_{\text{NH}} \in \mathbb{Z}$). Building upon local inequalities derived in the previous Section, the bound follows as:
\begin{equation}
    2\pi |C_\text{NH}|\leqslant \int_\text{BZ} \intd \left|F^{\text{RL}}_{12}\right| \leqslant\int_\text{BZ}\intd \Big( \left|Q^{\text{RL}}_{12}\right| + \left|Q^{\text{RL}}_{21}\right| \Big).
\end{equation}
We apply the arithmetic mean-geometric mean inequality to the right-hand side (RHS) of the QGT inequality Eq.~\eqref{QGT-inequality} to find:
\begin{align}
   2 \left|Q^{\text{RL}}_{12}\right| &\leqslant  ||\psi^{\text{R}}_n||^2\left(Q^{\text{RR}}_{11} + |Q^{\text{R}}_1|^2 \right)+||\psi^{\text{L}}_n||^2\left( Q^{\text{LL}}_{22} + |Q^{\text{L}}_2|^2 \right), \\
    2\left|Q^{\text{RL}}_{12}\right| &\leqslant  ||\psi^{\text{R}}_n||^2\left(Q^{\text{RR}}_{22} + |Q^{\text{R}}_2|^2 \right)+||\psi^{\text{L}}_n||^2\left( Q^{\text{LL}}_{11} + |Q^{\text{L}}_1|^2 \right).
\end{align}
Substituting in these inequalities, {we obtain the final bound on the non-Hermitian Chern number,
\begin{equation}
    2\pi|C|\leqslant \frac{1}{2}\int_\text{BZ} \intd \; \left\{||\psi_n^{\text{R}}||^2\left[\tr Q^{\text{RR}} + ||\mathbf{Q}^{\text{R}}||^2 \right] +  ||\psi_n^{\text{L}}||^2\left[\tr Q^{\text{LL}} + ||\mathbf{Q}^{\text{L}}||^2 \right]\right\}.
\end{equation}
}

\subsection{Gauge-invariant formulations of non-Hermitian quantum geometry}\label{Gauge-invariant-formulation}

\subsubsection*{Non-Hermitian projector formalism}
First, we establish some facts about the properties of projectors that will be needed later. Consider a non-orthonormal basis $\{\ket{\psi^{\text{R}}_n}\}$ and its dual basis $\{\ket{\psi^{\text{L}}_n}\}$. We will denote normalized vectors as $\ket{\tilde \psi} \equiv \ket{\psi}/||\psi||$. Furthermore, we note the identity 
\begin{equation}
    1 = \sum_n \ket{\psi^{\text{R}}_n}\bra{\psi^{\text{L}}_n} = \sum_n \ket{\psi^{\text{L}}_n}\bra{\psi^{\text{R}}_n}.
\end{equation}
Moreover, in two band systems
\begin{equation}
    1 = \ket{\tilde\psi^{\text{R}}_0}\bra{\tilde \psi^{\text{R}}_0} + \ket{\tilde \psi^{\text{L}}_1}\bra{\tilde \psi^{\text{L}}_1}.
\end{equation}

\subsubsection*{Gauge-invariant projector formalism for numerical computations}
We now employ a gauge-invariant formulation of non-Hermitian quantum geometry. We begin by noting that
\begin{equation}
    (\varepsilon_n -  H)\ket{\partial_\mu \psi^{\text{R}}_n} = (\partial_\mu  H - \partial_\mu\varepsilon_n)\ket{\psi^{\text{R}}_n}.
\end{equation}
Thus, for $n \neq m$ we have,
\begin{equation}
    \braket{\psi^{\text{L}}_m }{ \partial_\mu \psi^{\text{R}}_n} = \frac{\bra{\psi^{\text{L}}_m}\partial_\mu  H\ket{\psi^{\text{R}}_n}}{\varepsilon_n - \varepsilon_m},
\end{equation}
which follows from $\bra{\psi^{\text{L}}_n}H = \varepsilon_n \bra{\psi^{\text{L}}_n}$ and orthogonality of dual bases. This allows us to rewrite multiple equations in a gauge-independent way. First, consider the non-Abelian $\text{LR}$ QGT between bands $n, m$. Let $O$ be the set of occupied bands, then
\begin{align}
    Q^{\text{LR}}_{nm;\mu\nu} \equiv \bra{\partial_\mu \psi^{\text{L}}_n}(1-P_O^{\text{RL}})\ket{\partial_\nu \psi^{\text{R}}_m} 
    = \sum_{k\notin O} \braket{\partial_\mu \psi^{\text{L}}_n}{ \psi^{\text{R}}_k}\braket{\psi^{\text{L}}_k}{\partial_\nu \psi^{\text{R}}_m} 
    = \sum_{k \notin O}\frac{\bra{\psi^{\text{L}}_n}\partial_\mu  H\ket{\psi^{\text{R}}_k}\bra{\psi^{\text{L}}_k}\partial_\nu  H \ket{\psi^{\text{R}}_m}}{(\varepsilon_n - \varepsilon_k)(\varepsilon_m - \varepsilon_k)}.
\end{align}
Naturally, the same follows for the RL QGT elements, which can be obtained via:
\begin{align}
    Q^{\text{RL}}_{nm;\mu\nu} &= (Q^{\text{LR}}_{mn;\nu\mu})^* 
    =  \sum_{k \notin O}\frac{\bra{\psi^{\text{R}}_n}\partial_\mu  H^\dagger\ket{\psi^{\text{L}}_k}\bra{\psi^{\text{R}}_k}\partial_\nu  H^\dagger \ket{\psi^{\text{L}}_m}}{(\varepsilon_n - \varepsilon_k)^*(\varepsilon_m - \varepsilon_k)^*}.
\end{align}
We further turn to single band $\text{RR}$ and $\text{LL}$ QGTs. First, we consider the two-band case, which is particularly simple, namely,
\begin{align*}
     Q^{\text{RR}}_{ij} &= \bra{\partial_\mu \tilde \psi^{\text{R}}_0}(1-\ket{\tilde \psi^{\text{R}}_0}\bra{\tilde \psi^{\text{R}}_0})\ket{\partial_\nu \tilde \psi^{\text{R}}_0} = \braket{\partial_\mu \tilde \psi^{\text{R}}_0}{\tilde \psi^{\text{L}}_1}\braket{\tilde \psi^{\text{L}}_1}{\partial_\nu \tilde \psi^{\text{R}}_0} \\
    &= \left(\frac{\bra{\tilde \psi^{\text{L}}_1}\partial_\mu  H\ket{\tilde \psi^{\text{R}}_0}}{\varepsilon_0 - \varepsilon_1}\right)^* \frac{\bra{\tilde \psi^{\text{L}}_1}\partial_\nu {H}\ket{\psi^{\text{R}}_0}}{\varepsilon_0 - \varepsilon_1} 
    = \frac{\bra{\tilde \psi^{\text{R}}_0} \partial_\mu  H^\dagger \ket{\tilde \psi^{\text{L}}_1}\bra{\tilde \psi^{\text{L}}_1}\partial_\nu {H} \ket{\tilde \psi^{\text{R}}_0}}{|\varepsilon_0 -\varepsilon_1|^2}.
\end{align*}
Naturally, for the $\text{LL}$ QGT of the same band, we have identical expression via the mapping $\varepsilon \mapsto \varepsilon^*$, $ H\mapsto  H^\dagger$, $\text{L} \leftrightarrow \text{R}$:
\begin{equation}
     Q^{\text{LL}}_{\mu\nu} = \frac{\bra{\tilde \psi^{\text{L}}_0} \partial_\mu  H \ket{\tilde \psi^{\text{R}}_1}\bra{\tilde \psi^{\text{R}}_1}\partial_\nu  H^\dagger \ket{\tilde \psi^{\text{L}}_0}}{|\varepsilon_0 -\varepsilon_1|^2}.
\end{equation}
In the general case of a symmetric QGT of the $n^\text{th}$ band in a multiband system
\begin{align*}
      Q^{\text{RR}}_{n;\mu\nu} &= \bra{\partial_\mu \tilde \psi^{\text{R}}_n}(1-\ket{\tilde \psi^{\text{R}}_n}\bra{\tilde \psi^{\text{R}}_n})\ket{\partial_\nu \tilde \psi^{\text{R}}_n} = \sum_{m,k} \braket{\partial_\mu \tilde \psi^{\text{R}}_n}{\psi^{\text{L}}_m}\bra{\psi^{\text{R}}_m}(1-\ket{\tilde \psi^{\text{R}}_n}\bra{\tilde \psi^{\text{R}}_n})\ket{\psi^{\text{R}}_k}\braket{\psi^{\text{L}}_k}{\partial_\nu \tilde \psi^{\text{R}}_n} \\
     &= \sum_{m,k \neq n} \left(I_{mk} -\frac{I_{mn}I_{nk}}{I_{nn}}\right)\frac{\bra{\tilde \psi^{\text{R}}_n} \partial_\mu  H^\dagger\ket{\psi^{\text{L}}_m}\bra{\psi^{\text{L}}_k}\partial_\nu  H\ket{\tilde \psi^{\text{R}}_n}}{(\varepsilon_n - \varepsilon_m)^*(\varepsilon_n - \varepsilon_k)}
     ,
\end{align*}
where $I$ is the right basis overlap matrix: $I_{nm} = \braket{\psi^{\text{R}}_n}{\psi^{\text{R}}_m}$. Due to the normalization, the inverse matrix is the overlap matrix of the left eigenbasis~\cite{Silberstein:2020hi}. We note that in the last line, the $n=m=k$ term vanishes identically. This can naturally be extended to the $\text{LL}$ QGT:
\begin{equation}
     Q^{\text{LL}}_{n;\mu\nu}  = \sum_{m,k \neq n} \left(I^{-1}_{mk} -\frac{I^{-1}_{mn}I^{-1}_{nk}}{I^{-1}_{nn}}\right)\frac{\bra{\tilde \psi^{\text{L}}_n} \partial_\mu  H\ket{\psi^{\text{R}}_m}\bra{\psi^{\text{R}}_k}\partial_\nu  H^\dagger\ket{\tilde \psi^{\text{L}}_n}}{(\varepsilon_n - \varepsilon_m)(\varepsilon_n - \varepsilon_k)^*}.
\end{equation}
   
Finally, for the connection difference $Q$ we have
\begin{align}
    A^{\text{RR}}_{n;\mu} = \ii\frac{1}{I_{nn}} \sum_m\braket{\psi^{\text{R}}_n}{\psi^{\text{R}}_m}\braket{\psi^{\text{L}}_m}{\partial_\mu \psi^{\text{R}}_n} = \ii \sum_m\frac{I_{nm}}{I_{nn}} \frac{\bra{ \psi^{\text{L}}_m} \partial_\mu  H \ket{\psi^{\text{R}}_n}}{\varepsilon_n -\varepsilon_m}.
\end{align}
Once the relevant LR connection is subtracted, we obtain a correct formula for $Q^{\text{R}}$~\cite{Behrends2025}
\begin{equation}
    Q^{\text{R}}_{n;\mu} =  \ii\sum_{m\neq n} \frac{I_{nm}}{I_{nn}} \frac{\bra{\psi^{\text{L}}_m} \partial_\mu  H \ket{\psi^{\text{R}}_n}}{\varepsilon_n -\varepsilon_m},
\end{equation}
and for the $\text{L}$ connection differences:
\begin{equation}
    Q^{\text{L}}_{n;\mu} = \ii\sum_{m\neq n} \frac{I^{-1}_{nm}}{I^{-1}_{nn}} \frac{\bra{ \psi^{\text{R}}_m} \partial_\mu  H^\dagger \ket{\psi^{\text{L}}_n}}{\varepsilon_n^* -\varepsilon_m^*}.
\end{equation}
These formulae are sufficient to compute relevant quantities without numerical issues. Alternatively, a set of gauge-independent formulae may be developed by expressing the gauge-invariant objects in terms of generalized projectors and their derivatives; yet, we will not use this approach here.

\subsection{Optical weight inequalities in non-Hermitian systems}

In Hermitian systems, the QGT is closely connected to the conductivity (current-current response function)~\cite{Onishi:2024gl}. More precisely, the QGT is physically manifested in the optical weights~\cite{Kudinov1991}
\begin{equation}
    W^{\mu\nu} = \int_0^\infty\text{d}\omega\frac{\Re\sigma^\text{reg}_{\mu\nu}(\omega)}{\omega},
\end{equation}
where $\sigma_{\mu\nu}^\text{reg}$ is the regular part of conductivity (excluding the Drude peak $\propto \del_\mu\del_\nu\varepsilon$~\cite{Souza:2000cj}).
Notably, a similar connection exists in the case of wave-packet dynamics with non-Hermitian dynamics~\cite{Behrends2025}.
The regular part of the optical weight is related to the RR-QGT in $\mathcal{PT}$-symmetric systems. Here, we use this relationship to calculate a Chern bound in that case, as well as extend this relationship to arbitrary two-band systems.
 \\\\
 
Indeed, consider the position-current response. Generically, it is impossible to define a velocity operator in a generic non-Hermitian system~\cite{Behrends2025}, with the exception of quadratic open quantum systems, where we derive a local velocity operator. However, assuming minimal coupling to the vector potential, we have for the current operator~\cite{Behrends2025}
\begin{equation}
    {j}_\mu = \partial_{\mu} H.
\end{equation}
{{$\partial_{\mu} \equiv \partial_{k_\mu}$, consistently with the main text, and the Hamiltonian $H$ is a function of momentum, as e.g., the non-Hermitian Rice--Mele model.}
Using the non-Hermitian Kubo formula~\cite{Behrends2025} one can then derive the position-current response (i.e., conductivity) of a wave-packet centered at quasimomentum $\mathbf{k}$ in the $n^\text{th}$ band. The response is found to depend only on the time difference, allowing to restate it in the frequency domain as~\cite{Behrends2025}
\begin{align*}
    \sigma^{\mu\nu}(\omega,\mathbf{k}) =& \ii\left(\mathcal{P}\frac{1}{\omega}-\ii\pi\delta(\omega)\right)\partial_\mu\partial_\nu\varepsilon_{n\mathbf{k}} + \sum_{m\neq n} \left[\ii\frac{(\varepsilon_{n\mathbf{k}} - \varepsilon_{m\mathbf{k}})f^{nm}_{\mu\nu}(\mathbf{k})}{\varepsilon_{m\mathbf{k}}-\varepsilon_{n\mathbf{k}}-\omega} +\left(\ii\frac{(\varepsilon_{n\mathbf{k}} - \varepsilon_{m\mathbf{k}})f^{nm}_{\mu\nu}(\mathbf{k})}{\varepsilon_{m\mathbf{k}}-\varepsilon_{n\mathbf{k}}+\omega}\right)^*\right] \\
    &+\ii\omega\sum_{m\neq n}\left[ \frac{h^{nm}_{\mu\nu}(\mathbf{k})}{(\varepsilon_{m\mathbf{k}}-\varepsilon_{n\mathbf{k}}-\omega)^2}+\left(\frac{h^{nm}_{\mu\nu}(\mathbf{k})}{(\varepsilon_{m\mathbf{k}}-\varepsilon_{n\mathbf{k}}+\omega)^2}\right)^*\right],
\end{align*}
where
\begin{align}
    f^{nm}_{\mu\nu}(\mathbf{k}) &= \frac{I_{\mathbf{k}nm}}{2I_{\mathbf{k}nn}}\left[\braket{\psi^{\text{L}}_{\mathbf{k}m}}{\partial_{\nu}\psi^{\text{R}}_{\mathbf{k}n}}\left(\frac{\braket{\partial_\mu\psi^{\text{R}}_{\mathbf{k}n}}{\psi^{\text{R}}_{\mathbf{k}m}}-\braket{\psi^{\text{R}}_{\mathbf{k}n}}{\partial_\mu\psi^{\text{R}}_{\mathbf{k}m}}}{I_{\mathbf{k}nm}}-2\ii\Re A^{\text{RR}}_{\mu}\right)-\partial_\mu\braket{\psi^{\text{L}}_{\mathbf{k}m}}{\partial_\nu \psi^{\text{R}}_{\mathbf{k}n}}\right], \\
    h^{nm}_{\mu\nu}(\mathbf{k}) &= \frac{I_{\mathbf{k}nm}}{2I_{\mathbf{k}nn}}\braket{\psi^{\text{L}}_{\mathbf{k}m}}{\partial_\nu\psi^{\text{R}}_{\mathbf{k}n}}\partial_{\mu}(\varepsilon_{n\mathbf{k}} - \varepsilon_{m\mathbf{k}}).
\end{align}
Here, $I$ is the right basis overlap matrix: $I_{\mathbf{k}nm} = \braket{\psi^{\text{R}}_{\mathbf{k}n}}{\psi^{\text{R}}_{\mathbf{k}m}}$. In the following we will make use of the lemma:

\textit{Lemma.} For any compact $d$-dimensional manifold without boundary (e.g. $\mathbb{T}^d)$ $\mathcal{M}$ and a ($d-1$)-form $A$ derived from wavefunctions, which is gauge-invariant (such as anomalous Berry connection) the following holds:
\begin{equation}\label{eq:lemma}
    \int_\mathcal{M} \text{d}A = 0,
\end{equation}
where $\text{d}$ denotes an exterior derivative. Notably, this is not true for gauge-dependent forms such as Berry connection.

\textit{Proof.} Here we use the characterization of the integral of an exterior derivative as obstructions. This characterization was used to prove equality of $\text{RR}$, $\text{RL}$, $\text{LR}$ and $\text{LL}$ Chern numbers by~\cite{Shen2018}. Consider a gauge-invariant $(d-1)$-form. $A$, which arises from wavefunctions defined on $\mathcal{M}$. The only possibility for
\begin{equation}
    \int_{\mathcal{M}} \text{d}A \neq 0,
\end{equation}
is due to the impossibility of choosing the gauge of a wavefunction smoothly over the entire $\mathcal{M}$. We consider the case where two patches are sufficient for a gauge to be well-defined in each of them. Indeed, as $\mathcal{M}$ is compact, the number of patches is always finite, hence there are no issues with generalization. Then
\begin{equation}
    \int_{\mathcal{M}} \text{d}A = \int_{\partial P} A^I+\int_{\partial(\mathcal{M}\setminus P)}A^{II},
    \end{equation}
where $A^I, A^{II}$ are the forms defined in two different gauges (in patch $P$ and the rest of the $\mathcal{M}$). However, by assumption, $A$ is gauge-invariant, and by definition of outwards facing normal:
    \begin{equation}
        \partial P = -\partial(\mathcal{M}\setminus P),
    \end{equation}
hence,
\begin{equation}
    \int_{\mathcal{M}} \text{d}A = \int_{\partial P} (A^I-A^{II}) = 0.
\end{equation}

Then, we can for example consider the gauge-invariant one-form
\begin{equation}
    Q_\mu\, dk^{\bar \nu},
\end{equation}
where $\bar{1}\equiv2,\,\bar{2}\equiv1$, and $Q$ is the anomalous Berry connection. Thus, we find:
\begin{equation}
    0 = \int_\text{BZ}  \text{d}(Q_\mu\,  \text{d}k^{\bar \nu}) = \pm\int_\text{BZ} (\nabla \mathbf{Q})_{\mu\nu} \;  \text{d}k^x\wedge  \text{d}k^y.
\end{equation}
In order to compute the optical weight, we first compute two integrals:
\begin{align}
    \int_0^\infty \text{d}\omega\; \frac{1}{(\omega + z)^2} &= \frac{1}{z}, \\
    \int \text{d}\omega\;\frac{-z}{\omega(z -\omega)} &=-\int\text{d}\omega\; \left(\frac{1}{\omega}+\frac{1}{z-\omega}\right) = -\ln\omega+\ln(\omega-z),
\end{align}
where $z$ is a complex number with a nonvanishing imaginary part -- otherwise, a limiting procedure with an infinitesimal positive imaginary part would be necessary. The first integral applied to the $h$-dependent part of optical weight yields the expected result, while for the $f$-dependent part we use the second integral and note that for any function $F$:
\begin{equation}
    \int_\eta^\infty\frac{\text{d}\omega}{\omega}F(-\omega)^* = \left[\int_{-\infty}^{-\eta} \frac{\text{d}\omega}{\omega}F(\omega)\right]^*.
\end{equation}
Then, using the above results for
\begin{equation}
    F(\omega) = \ii\frac{(\varepsilon_{n\mathbf{k}} - \varepsilon_{m\mathbf{k}})f^{nm}_{\mu\nu}(\mathbf{k})}{\varepsilon_{m\mathbf{k}}-\varepsilon_{n\mathbf{k}}-\omega},
\end{equation}
we obtain the contribution to the optical weight,
\begin{equation}
    \ii f^{nm}_{\mu\nu}\left[\ln\left(\frac{\omega-z}{\omega}\right)\right]_\eta^\infty +\left\{\ii f^{nm}_{\mu\nu}\left[\ln\left(\frac{\omega-z}{\omega}\right)\right]_{-\eta}^{-\infty}\right\}^* = -\ii f^{nm}_{\mu\nu}\ln\left(\frac{-z}{\eta}\right) + \left[-\ii f^{nm}_{\mu\nu}\ln\left(\frac{z}{\eta}\right)\right]^*,
\end{equation}
where $z=\varepsilon_{m\mathbf{k}}-\varepsilon_{n\mathbf{k}}$ lies in the lower half-plane. Thus, the branch cut of the logarithm must be taken in the lower half-plane, yielding $\ln(-1) = \ii \pi$. Hence, we can rewrite the above equation as
\begin{equation}
    =\pi f^{nm}_{\mu\nu} + 2\Im\left[f^{nm}_{\mu\nu}\ln(\varepsilon_{m\mathbf{k}}-\varepsilon_{n\mathbf{k}})\right] - 2\Im\left[f^{nm}_{\mu\nu}\right]\ln\eta.
\end{equation}
Combining these two contributions yields
\begin{equation}
\begin{split}
    &\int_\eta^\infty\frac{\text{d}\omega}{\omega}\Re\sigma_{\mu\nu}^\text{reg}(\omega,\mathbf{k}) = 2\sum_{m\neq n}\left\{\Im\frac{h^{nm}_{\mu\nu}(\mathbf{k})}{\varepsilon_{m\mathbf{k}}-\varepsilon_{n\mathbf{k}}} -\left[\Im f^{nm}_{\mu\nu}(\mathbf{k})\right]\ln{\eta}+\Im\left[ f^{nm}_{\mu\nu}(\mathbf{k})\ln(\varepsilon_{m\mathbf{k}}-\varepsilon_{n\mathbf{k}})\right]+\frac{1}{2}\pi\Re f^{nm}_{\mu\nu}(\mathbf{k})\right\},
    \label{eq:Equalityf}
\end{split}
\end{equation}
where $\eta$ is the IR-cutoff, as the integral is generically logarithmically divergent. In what follows, we will assume that the $n^\text{th}$ band has the largest imaginary part of the energy, i.e. it is the fastest growing/slowest decaying. We will, without loss of generality, renumber the bands such that $n=0$.
\\\\
To derive the final form {for the two-band non-Hermitian Hamiltonians}, note that
    \begin{equation}\label{eq:HplusF}
    \begin{split}
    &2\Im\left\{\frac{h_{\mu\nu}^{nm}}{z} + f_{\mu\nu}^{nm}\ln{z}\right\} = \partial_\mu F_\nu +\Im\left\{\frac{I_{nm}}{I_{nn}}\braket{\psi^{\text{L}}_{m}}{\partial_\nu\psi^{\text{R}}_n}\ln{z}\left[\frac{\braket{\partial_\mu \psi^{\text{R}}_n}{\psi^{\text{R}}_{m}}}{I_{nm}}-i\left({A_{n\mu}^{\text{R}}}\right)^*\right]\right\}\\
    &=\partial_\mu F_\nu + \Im\left\{\frac{I_{nm}}{I_{nn}}\braket{\psi^{\text{L}}_{m}}{\partial_\nu \psi^{\text{R}}_n}\ln{z}\bra{\partial_\mu \psi^{\text{R}}_n}\left[\frac{\ket{\psi^{\text{R}}_{m}}}{\braket{\psi^{\text{R}}_n}{\psi^{\text{R}}_{m}}}-\frac{\ket{\psi^{\text{R}}_n}}{\braket{\psi^{\text{R}}_n}{\psi^{\text{R}}_n}}\right]\right\},
\end{split}
\end{equation}
with gauge-invariant
\begin{equation}
    F_\nu = -\Im\left\{\frac{I_{nm}}{I_{nn}}\braket{\psi^{\text{L}}_{m}}{\partial_\nu \psi^{\text{R}}_n}\ln{z}\right\},
\end{equation}
where $z$ is defined as before. Thus, we note that by the lemma on gauge-invariant differential forms, the integral of $\partial_\mu F_\mu$ over the Brillouin zone (BZ) vanishes. The remaining term is evaluated as follows: for two bands, it may be rewritten using  $\ket{\psi^{\text{R}}_1}\bra{\psi^{\text{L}}_1} = 1-\ket{\psi^{\text{R}}_0}\bra{\psi^{\text{L}}_0}$:
\begin{align*}
   & \sum_\mu\frac{I_{01}}{I_{00}}\braket{\psi^{\text{L}}_{1}}{\partial_\mu\psi^{\text{R}}_0}\ln{z}\bra{\partial_\mu \psi^{\text{R}}_0}\left[\frac{\ket{\psi^{\text{R}}_{1}}}{\braket{\psi^{\text{R}}_0}{\psi^{\text{R}}_{1}}}-\frac{\ket{\psi^{\text{R}}_0}}{\braket{\psi^{\text{R}}_0}{\psi^{\text{R}}_0}}\right] \\
    &=\sum_\mu\ln{z}\left[ \frac{\bra{\partial_\mu\psi^{\text{R}}_0}\Big(1-\ket{\psi^{\text{R}}_0}\bra{\psi^{\text{L}}_0}\Big)\ket{\partial_\mu\psi^{\text{R}}_0}}{\braket{\psi^{\text{R}}_0}{\psi^{\text{R}}_0}}-\braket{\partial_\mu \psi^{\text{R}}_0}{\psi^{\text{R}}_0}\frac{\bra{\psi^{\text{R}}_0}\Big(1-\ket{\psi^{\text{R}}_0}\bra{\psi^{\text{L}}_0}\Big)\ket{\partial_\mu\psi^{\text{R}}_0}}{\braket{\psi^{\text{R}}_0}{\psi^{\text{R}}_0}^2} \right] \\
    &=\ln{z}\tr Q^{\text{RR}}.
\end{align*}
Then, since {$Q^{\text{RR}}$ is Hermitian, $\tr Q^{\text{RR}}\in\mathbb{R}$}, we have,
\begin{equation}
    \Im(\ln{z}\tr  Q^{\text{RR}}) = \tr Q^{\text{RR}}\text{arg}(\varepsilon_1 -\varepsilon_0).
\end{equation}

\subsubsection*{Two-band inequalities}
We now turn to a generic two-band system, in particular, the trace of the optical weight. We first note that, upon integrating over the BZ and applying the lemma to the second term,
\begin{align}
     \int_\text{BZ} \intd\sum_\mu\sum_{n>0} f^{\mu\mu}_{0n} &= \int_\text{BZ}\intd \left(\tr Q^{\text{RR}} + \frac{\ii}{2}\nabla\cdot\mathbf{Q}^{\text{R}} \right) = \int_\text{BZ}\intd \tr Q^{\text{RR}},
\end{align}
where we have used the formula for $\sum_{m\neq n}f^{nm}$ derived in Ref.~\cite{Behrends2025}. Hence, since the trace of the symmetric QGT is real, we conclude that the logarithmically divergent contribution vanishes identically. For two-band systems, the remaining sums may be evaluated to yield,
\begin{equation}
    \sum_\mu W^{\mu\mu} = \int_\text{BZ}\intd\tr Q^{\text{RR}}\big[\pi+\arg(\varepsilon_{1\mathbf{k}}-\varepsilon_{0\mathbf{k}})\big].
\end{equation}
Since we are concerned with the response of the slowest decaying bands, as it is primarily responsible for the properties of the system, we have
\begin{equation}
    \arg(\varepsilon_{1\mathbf{k}}-\varepsilon_{0\mathbf{k}}) \in[-\pi,0].
\end{equation}
Thus, $\pi + \arg(\varepsilon_{1\mathbf{k}}-\varepsilon_{0\mathbf{k}})>0$, and we find a bound on the optical weight
\begin{equation}
    \sum_\mu W^{\mu\mu} \geqslant 2\pi\big[\pi+\min_{\mathbf{k}\in \text{BZ}}\arg(\varepsilon_{1\mathbf{k}}-\varepsilon_{0\mathbf{k}})\big]|C_\text{NH}| >0.
\end{equation}
This shows that the connection between optical weight and topology is not limited to systems with real spectrum, such as Hermitian or $\mathcal{PT}$-symmetric systems. It is worth noting that beyond the two-band limit, the sums cannot be explicitly evaluated, suggesting that the connection of conductivity to topology, if it exists, is much more convoluted.

{\subsubsection*{$N$-band generalizations}}

{
Further to the derivation of two-band inequalities, we consider a multiband generalization of the non-Hermitian optical weight conditions in $N$-band systems, with $N > 2$. We begin by generalizing the optical weight expression Eq.~\eqref{eq:Equalityf} to an $N$-band equality, upon setting $n=0$,

\begin{equation}
\begin{split}
    &\int_\eta^\infty\frac{\text{d}\omega}{\omega}\Re\sigma_{\mu\nu}^\text{reg}(\omega,\mathbf{k}) = 2\sum_{m\neq 0}\left\{\Im\frac{h^{0m}_{\mu\nu}(\mathbf{k})}{\varepsilon_{m\mathbf{k}}-\varepsilon_{0\mathbf{k}}} -\left[\Im f^{0m}_{\mu\nu}(\mathbf{k})\right]\ln{\eta}+\Im\left[ f^{0m}_{\mu\nu}(\mathbf{k})\ln(\varepsilon_{m\mathbf{k}}-\varepsilon_{0\mathbf{k}})\right]+\frac{1}{2}\pi\Re f^{0m}_{\mu\nu}(\mathbf{k})\right\}.
    \label{nh-conductivity}
\end{split}
\end{equation}
We further define $z_m \equiv \varepsilon_{m\textbf{k}} - \varepsilon_{0\textbf{k}}$, integrate over BZ, and take a trace over the $\mu, \nu = x,y$ indices, to obtain,
\begin{equation}
\begin{split}
     \sum_\mu W^{\mu\mu} = 2\sum_{\mu}\sum_{m\neq 0} \int_\text{BZ}\intd\left\{\Im\frac{h^{0m}_{\mu\mu}(\mathbf{k})}{z_m} -\left[\Im f^{0m}_{\mu\mu}(\mathbf{k})\right]\ln{\eta}+\Im\left[ f^{0m}_{\mu\mu}(\mathbf{k})\ln z_m\right]+\frac{1}{2}\pi\Re f^{0m}_{\mu\mu}(\mathbf{k})\right\}.
     \label{eq:NbandDecom}
\end{split}
\end{equation}
We further note that $\sum_{m>0} f^{0m}_{\mu \nu} = Q^{\text{RR}}_{\mu \nu} + \frac{\text{i}}{2} \partial_\mu Q_\nu$~\cite{Behrends2025}, thus, $\sum_{m>0} \int_\text{BZ}\intd \text{Im}~f^{0m}_{\mu \nu} = \int_\text{BZ}\intd \partial_\mu Q_\nu = 0$, using Eq.~\eqref{eq:lemma}, and hence the second term on the RHS vanishes,
\begin{equation}
\begin{split}
     \sum_\mu W^{\mu\mu} = 2\sum_{\mu}\sum_{m\neq 0} \int_\text{BZ}\intd\left\{\Im\frac{h^{0m}_{\mu\mu}(\mathbf{k})}{z_m} +\Im\left[ f^{0m}_{\mu\mu}(\mathbf{k})\ln z_m\right]+\frac{1}{2}\pi\Re f^{0m}_{\mu\mu}(\mathbf{k})\right\}.
    \label{nh-conductivity}
\end{split}
\end{equation}
We now utilize an $N$-band analog of the Eq.~\eqref{eq:HplusF},

    \begin{equation}
    \begin{split}
    &2\Im\left\{\frac{h_{\mu\nu}^{0m}}{{z_m}} + f_{\mu\nu}^{0m}\ln{z_m}\right\} = \partial_\mu F^m_\nu +\Im\left\{\frac{I_{0m}}{I_{00}}\braket{\psi^{\text{L}}_{m}}{\partial_\nu\psi^{\text{R}}_0}\ln{{z_m}}\left[\frac{\braket{\partial_\mu \psi^{\text{R}}_0}{\psi^{\text{R}}_{m}}}{I_{0m}}-{\text{i}}\left({A_{0\mu}^{\text{R}}}\right)^*\right]\right\}\\
    &=\partial_\mu F^{m}_\nu + \Im\left\{\frac{I_{0m}}{I_{00}}\braket{\psi^{\text{L}}_{m}}{\partial_\nu \psi^{\text{R}}_0}\ln{z_m}\bra{\partial_\mu \psi^{\text{R}}_0}\left[\frac{\ket{\psi^{\text{R}}_{m}}}{\braket{\psi^{\text{R}}_0}{\psi^{\text{R}}_{m}}}-\frac{\ket{\psi^{\text{R}}_0}}{\braket{\psi^{\text{R}}_0}{\psi^{\text{R}}_0}}\right]\right\},
\end{split}
\end{equation}
with an $N$-band generalization $F^m_\nu$ of a gauge-invariant tensor $F_\nu$,
\begin{equation}
    F^m_\nu = -\Im\left\{\frac{I_{0m}}{I_{00}}\braket{\psi^{\text{L}}_{m}}{\partial_\nu \psi^{\text{R}}_0}\ln{z_m}\right\}.
\end{equation}
We further introduce a shortcut notation for the second term ($\tau^m_{\mu \nu})$, which is also gauge-invariant,
\beq{}
\tau^m_{\mu \nu} = \frac{I_{0m}}{I_{00}}\braket{\psi^{\text{L}}_{m}}{\partial_\nu \psi^{\text{R}}_0} \bra{\partial_\mu \psi^{\text{R}}_0}\left[\frac{\ket{\psi^{\text{R}}_{m}}}{\braket{\psi^{\text{R}}_0}{\psi^{\text{R}}_{m}}}-\frac{\ket{\psi^{\text{R}}_0}}{\braket{\psi^{\text{R}}_0}{\psi^{\text{R}}_0}}\right] = \frac{1}{I_{00}} \bra{\partial_\mu \psi^{\text{R}}_0}  1 - P^{\text{RR}}_0 \ket{\psi^{\text{R}}_m} \bra{\psi^{\text{L}}_m} \ket{\partial_\nu \psi^{\text{R}}_0}.
\eeq
with $P^{\text{RR}}_0 \equiv \frac{\ket{\psi^{\text{R}}_0}\bra{\psi^{\text{R}}_0}}{\braket{\psi^{\text{R}}_0}{\psi^{\text{R}}_0}}$, and where $\text{ln}~z_m$ was factored out. Using Eq.~\eqref{eq:lemma}, $\int_\text{BZ}\intd \partial_\mu F^m_\nu = 0$. We further note that if the non-Hermitian Hamiltonian $H$ satisfies a commutation relation $[H, H^\dagger] = 0$, i.e., the self-energy matrix is band-diagonalizable, then: $\Im \tau^m_{\mu\mu}=0$ and $\Re \tau^m_{\mu\mu} \geq 0$, which we further use for the derivation of the generalized $N$-band non-Hermitian bounds on the optical weights. $[H, H^\dagger] = 0$ implies that the right basis is orthogonal, as the eigenbasis of the Hermitian part of $H$, and thus, upon gauge-fixing, it can be taken identical to the left basis, $\ket{\psi^{\text{R}}_i} = \ket{\psi^{\text{L}}_i}$. This in turns allows us to simplify the expression for~$\tau^m_{\mu\nu}$
\beq{}
    \tau^m_{\mu\mu} = \frac{1}{I_{00}I_{mm}}\left|\braket{\partial_\mu \psi^{\text{R}}_0}{\psi^{\text{R}}_m} \right|^2 \in \mathbb{R}^+\cup\{0\}.
\eeq 

Correspondingly, under these conditions: $\Im \tau^m_{\mu\mu}=0$ and $\Re \tau^m_{\mu\mu} \geq 0$, and the integral of the second term ($\tau^m_{\mu \nu}$) satisfies,
{\color{black}
\beq{}
    \sum_\mu\sum_{m > 0} \int_\text{BZ}\intd \text{Im}\{\tau^m_{\mu \mu}\ln z_m\} \geq \min_{m>0,\mathbf{k}\in \text{BZ}}\arg z_m \sum_\mu\sum_{m > 0} \int_\text{BZ}\intd \tau^m_{\mu \mu} = \min_{m>0,\mathbf{k}\in \text{BZ}}\arg z_m \sum_\mu \int_\text{BZ}\intd Q^{\text{RR}}_{\mu \mu}.
\eeq
}
By combining with the third term of Eq.~\eqref{eq:NbandDecom}, where $\sum_\mu \sum_{m>0} \int_\text{BZ}\intd ~\text{Re}~f^{0m}_{\mu \nu} = \int_\text{BZ}\intd~\text{tr}~Q^{\text{RR}}_{\mu \nu}$,
this culminates in an $N$-band inequality,
\begin{equation}
    \sum_\mu W^{\mu\mu} \geqslant \big[\pi+\min_{m>0,\mathbf{k}\in \text{BZ}}\arg z_m \big] \int_\text{BZ}\intd  \text{tr}~Q^{\text{RR}}.
\end{equation}
Finally, upon combining with the inequality: $\frac{1}{2\pi} \int_\text{BZ}\intd  \text{tr}~Q^{\text{RR}} \geq \frac{1}{2\pi} \int_\text{BZ}\intd  \text{tr}~|F^{\text{RL}}_{xy}| = |C_\text{NH}|$, and expanding $z_m$ in the non-Hermitian band eigenvalues $\varepsilon_{m\kv}$, $\varepsilon_{0\kv}$, we conclude with a generalized $N$-band non-Hermitian quantum geometric bound, 
\begin{equation}\label{eq:NbandFinal}
    \sum_\mu W^{\mu\mu} \geqslant 2\pi\big[\pi+\min_{m>0,\mathbf{k}\in \text{BZ}}\arg(\varepsilon_{m\mathbf{k}}-\varepsilon_{0\mathbf{k}})\big]|C_\text{NH}| >0.
\end{equation}
Here, we stress that the non-Hermitian Chern number $C_\text{NH}$ corresponds to the Chern number of the band $n=0$. By choice, $n=0$ corresponds to a slowest-decaying band with $\text{arg}~z_m \in [-\pi,0) < 0$. Therefore, Eq.~\eqref{eq:NbandFinal} captures the non-Hermitian quantum geometric bound on the optical response of the slowest-decaying band in an $N$-band system.
}\\

\section{Non-Hermitian geometric bounds in responses of open quantum systems}

In this Section, we explicitly construct non-Hermitian bounds in response functions from dissipative Lindblad equation~\cite{Daley:2014hc,Monkman2025}.
{We focus on response functions for vector operators: $\Sigma^K \propto \Sigma^{\text{R}}$, with $\Sigma^{\text{R}}  = \frac{H-H^\dagger}{2\text{i}}$ the anti-Hermitian part of the Hamiltonian.} 

As demonstrated in the main text, any quadratic non-Hermitian Hamiltonian may be obtained from a Lindbladian, which contains only jump operators of the form
\begin{equation}
    J_m = \vec{a}\cdot \vec{c}.
\end{equation}
{\color{black} Consistent with the main text, below we focus on the case of $[{H},\Sigma^{\text{R}}]=0$.}
Here, we will derive geometric bounds for open quantum system, where the coupling to the bath is purely absorptive. To that end, we make use of Ref.~\cite{Talkington:2024gf} results for the case of $\Delta=0$, upon setting either $\vec{a}=0$ or $\vec{b}=0$. In these cases, for the self energies, we have:
\begin{align}
    \Sigma^K = \pm 2\ii \Sigma^{\text{R}}.
\end{align}
We now derive a geometric bound, which holds if $[{H},\Sigma^{\text{R}}]=0$. In this case, the eigenbasis of the non-Hermitian Hamiltonian is the same as that of its Hermitian part. Thus the form of the open-system response function for a vector operator is:
\begin{equation}
    \Pi_{ij}(\om) = \lb \pi_{ij}(\om)+\pi_{ij}(-\om)^*\rb,
\end{equation}
where we have defined a polarization bubble:
\begin{equation}
    \pi_{ij}(\om) \equiv \pm\int_{-\infty}^\infty \frac{\dd\om'}{2\pi} \Tr \lb O^i G^{\text{R}}(\om')O^j G^{\text{R}}(\om+\om')\Sigma^{\text{R}} G^A(\om+\om')\rb.
\end{equation}
Notably, the consequence of $[H,\Sigma^{\text{R}}]=0$ is that $G^{\text{R}},G^A,G^K,\Sigma^{\text{R}},\Sigma^K$ are all simultaneously diagonalizable, and the basis which does so is orthonormal. We employ it in the formula for $\pi_{ij}$ to arrive at:
\begin{equation}
    \pi_{ij}(\om) = \pm\sum_{nm}\int_{-\infty}^\infty\frac{\dd\om'}{2\pi} O^i_{nm}G^{\text{R}}(\om')_{m}O^j_{mn}G^{\text{R}}_n(\om+\om')\Sigma^{\text{R}}_nG^A_n(\om+\om').
\end{equation}
This integral may be evaluated by closing the contour in the upper half plane, where only the advanced Green's functions have poles. Thus, we arrive at:
\begin{align*}
    \pi_{ij}(\om) &= \pm\ii\sum_{nm}O^i_{nm}G^{\text{R}}_m(\varepsilon_n^*-\om)O^j_{mn}G^{\text{R}}_n(\varepsilon_n^*)\Sigma^{\text{R}}_n = \pm\ii\sum_{nm}O_{nm}^i\lb O_{nm}^j\rb^*\frac{1}{\varepsilon_n^*-\varepsilon_m-\om}\frac{\Im\varepsilon_n}{\varepsilon^*_n-\varepsilon_n} \\
    &= \mp\frac{1}{2}\sum_{nm}O^i_{nm}O^j_{mn} \frac{1}{E_{nm} - \om +\ii\Sigma_{nm}''} = \mp\frac{1}{2}\sum_{nm}O^i_{nm}O^j_{mn} \frac{E_{nm} - \om -\ii\Sigma_{nm}''}{(E_{nm} - \om)^2 +\lb\Sigma_{nm}''\rb^2},
\end{align*}
where $\epsilon,\,\Sigma''$ are the real and imaginary parts of the complex energy of the non-Hermitian effective Hamiltonian, and:
\begin{align}
    \Sigma_m''&\geqslant0, \\
    E_{nm} &= E_n-E_m, \\
    \Sigma_{nm}'' &= \Sigma_n'' + \Sigma_m''.
\end{align}
 We can combine these results to find that for the absorptive part of the response function:

\begin{align}
    \pi_{ij}^\text{abs}(\om) = \frac{1}{2\ii}\lb \pi_{ij}(\om) - \pi_{ji}(\om)^* \rb = \mp\frac{1}{4\ii}\sum_{nm} O_{nm}^iO_{mn}^j \lb \frac{E_{nm} - \om -\ii\Sigma_{nm}''}{(E_{nm} - \om)^2 +\lb\Sigma_{nm}''\rb^2} - \frac{E_{nm} - \om +\ii\Sigma_{nm}''}{(E_{nm} - \om)^2 +\lb\Sigma_{nm}''\rb^2} \rb \\
    =\pm\frac{1}{2}\sum_{nm} O_{nm}^iO_{mn}^j \frac{\Sigma_{nm}''}{(E_{nm} - \om)^2 +\lb\Sigma_{nm}''\rb^2} = \pm \frac{\pi}{2} \sum_{nm}L_{nm}(\om) O_{nm}^iO_{mn}^j,
\end{align}
where the Lorentzian $L_{nm}(\omega)$ is defined as in the main text, with $\forall_n:\Sigma'_n=0$.

\bibliographystyle{apsrev4-1}
\bibliography{references.bib}